\documentclass[HARVARD,Times2COL]{WileyNJDv5} %STIX1COL,STIX2COL,STIXSMALL

\usepackage{siunitx}
\usepackage{placeins}
\usepackage{graphicx}
\usepackage[space]{grffile}
\usepackage{latexsym}
\usepackage{textcomp}
\usepackage{longtable}
\usepackage{tabulary}
\usepackage{booktabs,array,multirow}
\usepackage{amsfonts,amsmath,amssymb}
\usepackage{url}
\usepackage{hyperref}
\hypersetup{colorlinks=false,pdfborder={0 0 0}}
\usepackage{etoolbox}
\usepackage{fix-cm}

\articletype{Research Article}%

\received{Date Month Year}
\revised{Date Month Year}
\accepted{Date Month Year}
\journal{Journal}
\volume{00}
\copyyear{2024}
\startpage{1}

\begin{document}

\title{Differential reflectivity columns and hail – linking C-band radar-based estimated column characteristics to crowdsourced hail observations in Switzerland}

\author[1,2]{Martin Aregger}
\author[1,2]{Olivia Martius}
\author[3]{Urs Germann}
\author[3]{Alessandro Hering}

\authormark{AREGGER \textsc{et al.}}
\titlemark{Differential reflectivity columns and hail – linking C-band radar-based estimated column characteristics to crowdsourced hail observations in Switzerland}

\address[1]{\orgdiv{Institute of Geography}, \orgname{University of Bern}, \orgaddress{\state{Bern}, \country{Switzerland}}}

\address[2]{\orgdiv{Oeschger Centre for Climate Change Research}, \orgname{University of Bern}, \orgaddress{\state{Bern}, \country{Switzerland}}}

\address[3]{\orgdiv{Division for Radar, Satellite and Nowcasting}, \orgname{MeteoSwiss}, \orgaddress{\state{Locarno-Monti}, \country{Switzerland}}}

\corres{Corresponding author Martin Aregger, Institute of Geography (GIUB), University of Bern, Hallerstrasse 12, 3012 Bern, Switzerland \email{martin.aregger@unibe.ch}}

\fundingInfo{This work is part of the scClim project (project no. CRSII5\_201792) funded by the Swiss National Science Foundation.}

\abstract[Abstract]{Differential reflectivity columns (${Z_{DR}C}$) have been shown to provide information about a storm’s updraft intensity and size. The updraft’s characteristics, in turn, influence a severe storm’s propensity to produce hail and the size of said hail. Consequently, there is the potential to use ${Z_{DR}C}$ for the detection and sizing of hail. In this observational study, we investigate the characteristics of ${Z_{DR}C}$ (volume, height, area, maximum $Z_{DR}$ within) automatically detected on an operational C-band radar network in Switzerland and relate them to hail on the ground using 173’000 crowdsourced hail reports collected over a period of 3.5 years.

We implement an adapted version of an established ${Z_{DR}C}$ detection algorithm on a 3D composite of ${Z_{DR}}$ data derived from five Swiss weather radars. The composite, in combination with the dense network of radars located on differing altitudes up to 3000 m.a.s.l, helps to counteract the effects of the complex topography of the study region. The alpine region presents visibility and data quality challenges, which are especially crucial for measuring ${Z_{DR}C}$.

Our analysis finds ${Z_{DR}C}$ present in most hail-producing storms, with higher frequencies in storms producing severe hail. Further, when looking at lifetime maximum values, we find significant differences in various ${Z_{DR}C}$ characteristics between hail-producing and non-hail-producing storms. We also attempt to determine thresholds to differentiate between storm types. 

The temporal evolution of the ${Z_{DR}C}$ proves challenging to investigate due to their intermittent nature. Nevertheless, the peak values of the ${Z_{DR}C}$ characteristics are most often measured 5-10 minutes before the first hail reports on the ground, highlighting the potential for ${Z_{DR}C}$ to be used in warning applications.}


%\jnlcitation{\cname{%
%\author{Aregger M.},
%\author{Martius O.},
%\author{Germann U.}, and
%\author{Hering A.}}.
%\ctitle{On simplifying ‘incremental remap’-based transport schemes.} \cjournal{\it J Comput Phys.} %\cvol{2021;00(00):1--18}.}

\maketitle
\section{Introduction}

Hail causes significant economic damage worldwide \citep{pucik.etal_2019,hoeppe_2016,allen.etal_2020}. For example, in 2021, Switzerland experienced a series of extensive and severe hailstorms causing significant damage to property and crops, with several insurance companies reporting record damages \citep{kopp.etal_2023a}. Hail damages can occur at varying hail sizes. Small hail (i.e., diameter < 2 cm), can severely impact crops, especially when combined with a high density of hailfall or coinciding with strong winds \citep{changnon_1999, sanchez.etal_1996}. However, even small hail can impact infrastructure when it causes clogging of drainage, leading to compounding flood events \citep{friedrich.etal_2019}. Hail larger than 2 cm can dent vehicle bodies \citep{hohl.etal_2002}, and diameters larger than 2.5 cm are typically required for damages to roofs and blinds to occur, depending on the construction standard \citep{stucki.egli_2007, hagelregister_2024}. At hailstone sizes exceeding approximately 4 cm, destroyed roof tiles and broken car windshields can be expected \citep{heymsfield.wright_2014,pucik.etal_2019}. It follows that relatively small changes in hail size (i.e. +- 1cm) can have vastly different consequences in terms of damage, which has implications for the accuracy required of any hail size measurement or forecasting product.

Ground observations of hail are the gold standard of hail size measurement in terms of accuracy. Switzerland has a network of automated hail sensors that provide ground observations \citep{kopp.etal_2023} However, it is currently limited to 80 stations in three small regions; hence, for much of the country, no direct ground observations are available.  The rarity and small spatial extent of hail swaths make it difficult to acquire long-term ground-based hail datasets with extensive spatial coverage \cite[e.g.][]{kopp.etal_2023a}. Radar-based approaches are widely used to create spatially and temporally highly resolved hail datasets \cite[e.g.][]{nisi.etal_2016,fluck.etal_2021}. Various relationships between radar variables and hail have been used to estimate hail occurrence and size \citep{allen.etal_2020}. The radar-based hail algorithms are tuned using (limited) ground-truth data \cite[e.g.][]{treloar_1998, foote.etal_2005,waldvogel.etal_1979,joe.etal_2004,murillo.homeyer_2019,witt.etal_1998}.

In Switzerland, the single-polarisation radar-based algorithms, Probability of Hail (POH) \citep{waldvogel.etal_1979, foote.etal_2005} for hail detection and Maximum Expected Severe Hail Size (MESHS) \citep{treloar_1998, joe.etal_2004} for hail size estimation, have been in operational use for more than 15 years. Various studies have used both algorithms extensively to characterise hail in Switzerland \cite[e.g.][]{nisi.etal_2016,nisi.etal_2018, schroeer.etal_2023, kopp.etal_2024}. A POH threshold of 80\% has been used as an indicator for hail presence for hail \citep{nisi.etal_2016,barras.etal_2019}. MESHS has also been used with some success to differentiate between smaller and larger hail \citep{nisi.etal_2016,barras.etal_2019,kopp.etal_2023a}. However, \citet{schmid.etal_2024} recently developed hail damage impact functions for vehicles and buildings based on an extensive set of insurance data using MESHS for calibration. Their hail damage estimates showed considerable uncertainties, which they attributed to the limitations of MESHS. In another attempt, they created damage curves based on crowdsourced hail reports, substantially improving the spatial representation of large hail compared to MESHS, indicating potential for improvement.

Since the development and implementation of MESHS, the Swiss radar network has been upgraded from single-polarisation to dual-polarisation C-band doppler radars; additionally, two new radars in high-altitude locations were added to improve the observation coverage in the complex topography of the Alps \citep{germann.etal_2022}. This new radar network, in combination with a uniquely large dataset of  more than 173'000 crowdsourced surface observations \citep{barras.etal_2019} provides a promising background for investigating new approaches for dual-polarisation radar-based hail-size estimation.

In the literature, different categories of radar-based hail detection techniques can be found. The techniques can be split into direct hail detection techniques, detection of indirect signatures caused by hail, and detection of indicators of storm intensity related to hail \cite[and sources cited therein]{allen.etal_2020}. Direct hail detection has proven difficult. 

\citet{ryzhkov.etal_2013,ryzhkov.etal_2013b} and \citet{ortega.etal_2016} have introduced a fuzzy logic-based Hail Size Discrimination Algorithm (HSDA) for S-band radars in the United States, which achieved a POD of 0.594 when evaluated against roughly 2000 hail reports submitted for 79 thunderstorms. \citet{schmidt_2020} attempted multiple approaches to adapt the HSDA to work with C-band radar data. However, they could not adjust the algorithm sufficiently to become a “usable tool to detect hail and discriminate its size”.

The indirect detection of hail uses radar signatures related to hail. A frequently studied signature is the “Three-Body Scattering Signature” (TBSS), which is linked to large hail; however, studies currently show an ambiguous relationship between TBSS and hail size and more research is needed \citep{allen.etal_2020}. \citet{rigo.farnell_2023} recently analysed TBSS signatures on hail days in Catalonia using C-band single polarisation radars. Their preliminary analysis is promising as they found TBSS signatures in 96\% of days with large hail (>= 2cm) and in 72\% of the days with small hail (< 2cm).

The final group of hail-size estimation techniques uses storm intensity-related signatures. Various approaches have been tested \cite[and sources therein]{allen.etal_2020}. MESHS is such a technique. It uses a relationship between the freezing level height and the 50 dBZ maximum echo height, which relates to storm intensity, to determine hail size \citep{treloar_1998, joe.etal_2004}. Similarly, the frequently used MESH \citep[Maximum Estimated Size of Hail,][]{witt.etal_1998} uses a temperature-weighted vertical integration of reflectivity to estimate hail size, which also depends on storm intensity.

Other approaches are based on determining the characteristics of the updraft in the convective storm. A key ingredient for large hail to form is the time the hailstones spend in a storm area conducive to growth \citep{allen.etal_2020}. The growing time is dictated by the hailstone's trajectory through the storm and depends on the updraft speed in relation to the hailstone’s fall speed and the updraft width. The wider the updraft, the larger the volume of air in which hailstone growth-relevant micro-physical processes can occur and the larger the potential hailstone embryo source region \citep{nelson_1983,dennis.kumjian_2017}. 

One approach to identify hail from polarimetric radar information are so-called differential reflectivity columns ($Z_{DR}C$). Differential reflectivity ($Z_{DR}$) describes the ratio between the reflectivities measured in the horizontal and the vertical polarisation. It can be interpreted as a measure of the shape of the target. For raindrops, $Z_{DR}$ is positive, and $Z_{DR}$ values become higher the larger the raindrops because the oblateness of raindrops increases with size due to aerodynamic drag \citep[e.g.][]{fabry_2015}. 

Early polarimetric observations of storms have detected vertically contiguous structures of elevated $Z_{DR}$ values within, which were collocated with the updrafts \cite[e.g.][]{illingworth.etal_1987,conway.zrnic_1993,ryzhkov.etal_1994,brandes.etal_1995,kumjian.etal_2010}. These vertically contiguous areas of elevated $Z_{DR}$ above the freezing level were coined with the name $Z_{DR}C$. It was concluded that they are a signature of large liquid drops lifted above the 0°C level by the updrafts within the convective cells \citep{illingworth.etal_1987}. 

More recent modelling studies \citep[e.g.][]{kumjian.etal_2014,snyder.etal_2015,ilotoviz.etal_2018,carlin.etal_2017} have clarified how the $Z_{DR}C$ form and how the $Z_{DR}C$ life cycle relates to storm characteristics such as the updraft intensity. \citet{kumjian.etal_2014} used a modelling approach to confirm that $Z_{DR}C$ represent large raindrops lofted above the freezing level by the updraft, as was hypothesised in observational studies. They further elaborate on the micro-physical characteristics leading to the formation and decay of the $Z_{DR}C$ (including the composition of hydrometeors influencing the $Z_{DR}C$). They conclude that $Z_{DR}C$ may be used as an indicator of storm strength and severity, and additionally, they also found evidence for the potential of $Z_{DR}C$ to be used for nowcasting hail at the ground.

Since then, a multitude of studies have set out to detect $Z_{DR}C$ manually and automatically and to use them as a prognostic tool for severe storms with hail, supercells, and tornadoes \citep{snyder.etal_2015,vandenbroeke_2021,plummer.etal_2018,kuster.etal_2019,segall.etal_2022,schmidt_2020,french.kingfield_2021,wilson.broeke_2022, broeke_2017, krause.klaus_2024,lo.etal_2024} These studies found that various $Z_{DR}C$ characteristics, such as the $Z_{DR}C$ height, the areal extent, and the maximum $Z_{DR}$ value within the $Z_{DR}C$ can be linked to storm severity. 

Most studies focus on convection in the US observed by S-band radars. Less is known about $Z_{DR}C$ in C-band radar observations, especially not in complex orography such as the Swiss Alps. C-band radars offer better weather-to-clutter ratios compared to S-band which is an especially important aspect in mountainous terrain \citep{germann.joss_2004}. Studies using both wavelengths have shown that there are pronounced differences in the radar signatures, especially for melting hail, which makes it clear that $Z_{DR}C$ detection algorithms developed for S-band radars must be adapted to be used with C-band radar observations \citep{picca.ryzhkov_2012,kaltenboeck.ryzhkov_2013}. Further, many studies are limited to relatively small observational datasets, especially for hail. 

In this study we first propose an implementation of a $Z_{DR}C$ detection algorithm on radar observations from the network of five operational C-band doppler radars in Switzerland. We then investigate the characteristics of $Z_{DR}C$ in the complex topography of Switzerland and use a large dataset of crowdsourced hail ground reports to show how these characteristics relate to hail. Specifically, we aim to use the detected $Z_{DR}C$ as a proxy for updrafts to answer the following questions:

\begin{itemize}
    \item Do the characteristics of $Z_{DR}C$ relate to observed hail on the ground?
    \item Can these $Z_{DR}C$ characteristics be used to estimate the observed hail size on the ground?
    \item Can the $Z_{DR}C$ be used to nowcast hail at the ground?
\end{itemize}  

In the following chapters, we introduce the radar (section 2.1) and crowdsourced data (section 2.2), followed by a small summary of $Z_{DR}C$ detection algorithms and an explanation of our own implementation (section 3.1). Then, we present our approach to combining the different datasets (sections 3.1 and 3.2). The results of how different $Z_{DR}C$ characteristics link to hail at the ground are shown in section 4.1, followed by our observations regarding nowcasting potential (section 4.2). Finally, the results are discussed (section 5), and general conclusions are drawn (section 6).

\section{Data}

To detect the $Z_{DR}C$, we use operational radar data (section 2.1) from the Rad4Alp network operated by MeteoSwiss \citep{germann.etal_2022} in combination with analysis data from the numerical weather prediction model COSMO-1E \citep{steppeler.etal_2003,klasa.etal_2018,klasa.etal_2019}. As “ground-truth” data for hail occurrence and size, we use crowdsourced hail reports, which were submitted by the users of the MeteoSwiss weather application (Chapter 2.2).

\begin{figure*}[th!]
\begin{center}
\includegraphics[width=1.5\columnwidth]{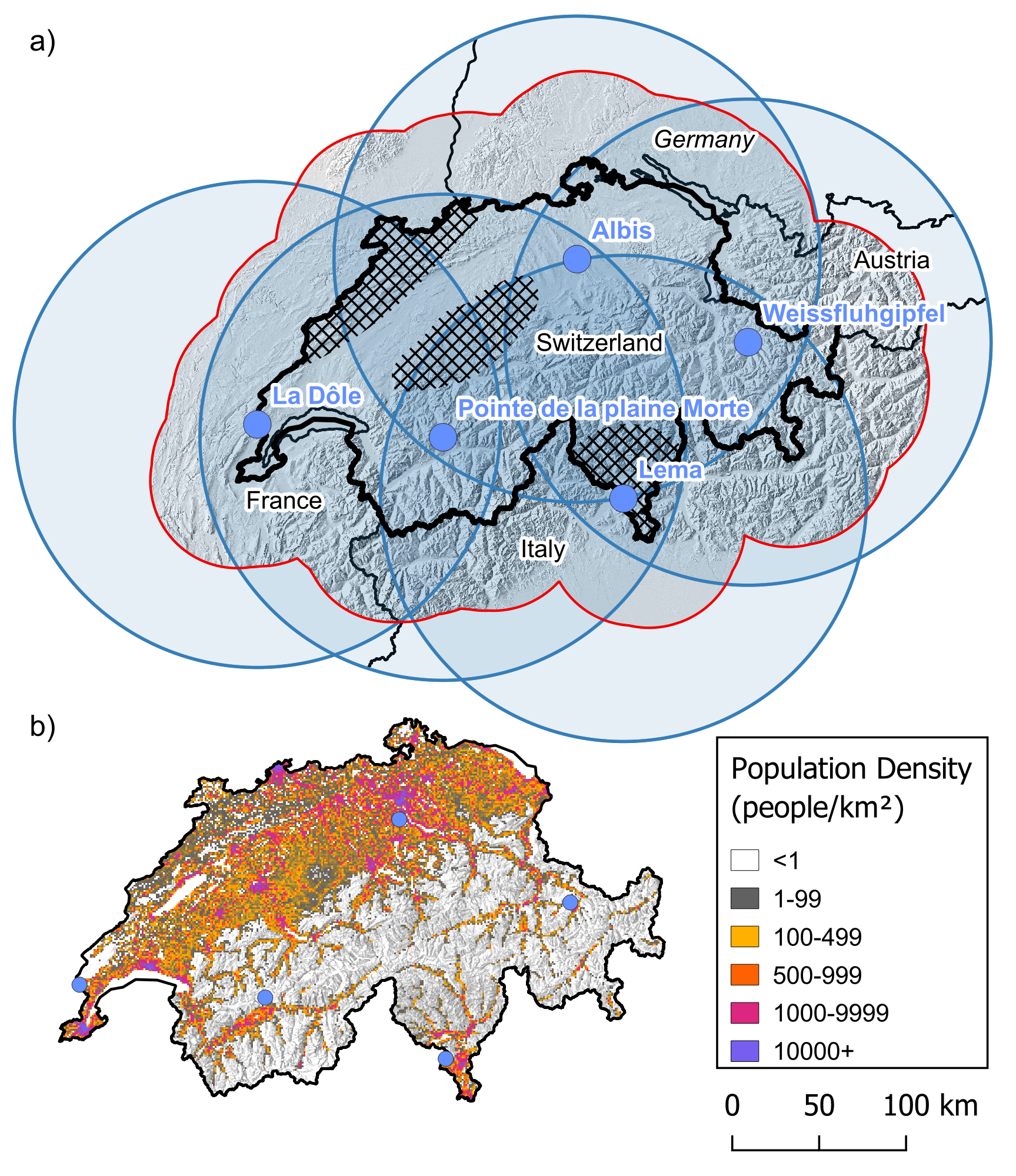}
\caption{{a) The study area (red outline) and its topography consisting of Switzerland (bold black outline) and a 50 km buffer around its borders. Further, the locations of the five radars of the Swiss weather radar network are indicated with blue dots and named, and the radar coverage up to a range of 140 km is shown with blue outlines. The particularly hail-prone areas (Jura, northern and southern Prealps) are marked with black hatching. b) The population density of Switzerland at a resolution of 1x1 km.
{\label{study area}}%
}}
\end{center}
\end{figure*}

The data are analysed for all months between January 2020 and September 2023. The study area contains Switzerland with a 50 km buffer zone around its borders, as shown in Fig. \ref{study area}a. The area is determined by the availability of crowdsourced reports, which are provided mainly by the population of Switzerland and people living close to its border.

Additionally, high-resolution population density data \citep{bfs_2021} is included in the analysis (Fig. \ref{study area}b). It is used to determine potentially non-hail-producing convective storm cells (chapter 3).

\subsection{Radar Data}

The Swiss operational weather radar network comprises five polarimetric C-band Doppler radars (see Fig. \ref{study area}a). The radars are located at altitudes between 900 to 3000 m asl to provide good coverage in Switzerland's complex topography, which includes the Jura Mountains, the Prealps, and the Alps. Using an interleaved scanning strategy, each radar provides a high vertical resolution by sampling 20 elevation scans at angles from -0.2 to 40° every 5 minutes \citep[][Fig. 14]{germann.etal_2022}, with the observational range extending up to 246 km. The radars sample the polarimetric moments at an angular 1° and an 83 m radial resolution. 

We use the data after pre-processing, which includes clutter cancellation and radial integration to 500m gates \citep[][Fig. 15 and 18]{germann.etal_2022}. All 20 elevation scans for $Z_{DR}$ and the Correlation Coefficient ($\rho_{HV}$) are extracted for each radar and used as input for the $Z_{DR}C$ detection algorithm every five minutes. 

\citet{friedrich.etal_2009} looked at the impact of clutter on $Z_{DR}$ measurements from C-band radars. They found that the reflectivity of precipitation needs to be, on average, 5.5 $dB$ higher than the reflectivity of ground clutter to achieve a precision of 0.2 $dB$ for $Z_{DR}$ measurements. In this context, it is important to underline that a comprehensive strategy is implemented for offline and online calibration and monitoring of the radars. More specifically, the operational weather radars' horizontal and vertical polarisation channels are automatically calibrated with a noise source every 2.5 minutes. The monitoring/calibration strategy is presented in \citet[][section 4.8]{germann.etal_2022}. In addition, the stability of the two channels is automatically monitored during the daytime using measurements of the sun. See for instance \citet{gabella.etal_2016}. 

The radar scan program plays a crucial role in identifying ZDR columns. It is essential to have many sweeps with small increments in the elevation angle to obtain a sufficiently high resolution in the vertical dimension. At the same time, it is also important to have high temporal resolution, as thunderstorms evolve rapidly in time. The scan program of the Swiss radars with 20 sweeps repeated every 5 minutes is particularly suited for this task.  
The analysis domain for this study falls within the 140km range of the radars where radar data quality and resolution are high.  
Further, a Cartesian maximum reflectivity composite product (MaxEcho) derived from all five radars is used, which is operationally produced every five minutes with a grid resolution of 1 km x 1 km. A discussion of the data quality of the radar observations can be found in \citet{feldmann.etal_2021}.

Additionally, to assess the potential of $Z_{DR}C$  for hail detection and size classification, we compare it to the currently operational single-polarisation radar-based hail products MESHS and POH. Both products are computed on a 1x1 km resolution every 5 minutes and are based on a relationship between the 0°C-isotherm height ($H_0$) from the numerical weather prediction model COSMO-1E and the heights of the highest elevations at which a $45 \, dBZ$ (POH) or $50 \, dBZ$ (MESHS) echo was detected (EchoTop). For details on their computation and a comprehensive verification, refer to \citet{trefalt.etal_2022} and \citet{kopp.etal_2024}.

\subsection{Crowdsourced Hail Reports}

Since May 2015, users of the MeteoSwiss weather app have the option to report hail occurrence and hail size. The reporting function asks users to choose from a selection of hail-size categories and records the time and location of the report. The categories are based on comparisons to commonplace objects such as “coffee bean”, “five francs coin” or “golf ball”. Throughout the years, the reporting function has undergone multiple changes; a new category was added, category descriptions were adapted to include different size estimates in mm, and for a time period, the function was placed less prominently. Consequently, depending on which app version a user has, their reports might differ slightly. For this reason, we use an aggregated version of the categories here instead of the seven currently available categories. We separate hail into small (ca. <=10mm, this category might also contain graupel), medium (ca. 15-35mm) and large (ca. >35mm) hailstones. The hail reports also undergo a number of plausibility filtering steps, as described in \citet{barras.etal_2019} and \citet{kopp.etal_2024}. Further, the reports are also filtered based on radar reflectivity. For a report to be valid, there must be a minimum radar reflectivity of 35 dBZ measured within a radius of 4 km from the report in the time frame from 15 minutes before to 15 minutes after the report. After filtering, the database contains approximately 173’000 reports for the study period (January 2020 - September 2023). 129’000 in the small category, 38’000 in the medium category and 6000 in the large category. %Despite the filtering, there can still be errors in the reports \citep{kopp.etal_2024} and we therefore require at least three consistent reports for a confirmed hail sighting. 

\subsection{TRT}
Convective storm objects are identified using the Thunderstorms Radar Tracking (TRT) algorithm \citep{hering.etal_2004, hering.etal_2008}. TRT uses the 2D Cartesian maximum reflectivity field to detect storms with a dynamic thresholding approach with minimum thresholds varying between 36-48 dBZ. It can simultaneously identify and track storms at different life cycle stages. The tracking is based on the geographical overlapping of cells in successive 5-minute time steps, considering their weighted average motion over the past three detections. To remove spurious detections of small, short-lived cells, we discard all cells with a lifetime of less than 30 minutes. Further, we only consider cells which spend their entire lifetime within the study area. The final dataset consists of 54'624 TRT-cells detected during the study period.

\section{Methods}

\subsection{Differential Reflectivity Column Detection}

In this study, we use a large sample of convective storms for which we identify the associated $Z_{DR}C$ throughout their lifetime. This requires a robust automated way to detect the $Z_{DR}C$. Further, for operational forecasters, the manual identification of $Z_{DR}C$ may be too time-consuming and automated detection may increase the potential usability of $Z_{DR}C$ in nowcasting \citep{kuster.etal_2020}.

One of the first automated $Z_{DR}C$ detection algorithms was published by \citet{snyder.etal_2015} (SN15 from here on). Their algorithm was implemented in the US on level-2 data from single radars of the WSR-88D radar network and consists of the following steps: For pre-processing, the $Z_{DR}$ Data are filtered with a 5-range-gate-moving average filter to reduce noisiness and then interpolated onto a 3D-grid (0.0025° lon x 0.0025° lat x 250 m height). Then, the number of vertically consecutive grid points with $Z_{DR} \geq1dB$ above the 0°C level (from hourly Rapid Refresh analysis data) is counted at each horizontal grid point to produce a 2D-$Z_{DR}C$-height product. This product is then smoothed further using a Gaussian filter. To remove potential TBSS contamination of the detected $Z_{DR}C$, all radar gates with $\rho_{HV}$ < 0.8 are filtered out. The SN15 has been used in multiple studies in the US and other countries with several adaptions \cite[e.g.][]{kuster.etal_2019,plummer.etal_2018,schmidt_2020,reinoso-rondinel.etal_2021, lo.etal_2024}

SN15 focused on the height of the $Z_{DR}C$ above the freezing level. Follow-up studies in the US investigated $Z_{DR}C$ characteristics associated with tornadic supercells and they identified the $Z_{DR}C$ area, specifically at 1 km height above the freezing level, as an important variable \citep{broeke_2016,wilson.broeke_2022,segall.etal_2022,broeke.etal_2023}. 

A novel approach for $Z_{DR}C$ area detection was introduced by \citet{french.kingfield_2021}. Their approach is based on the TRENDSS algorithm introduced in earlier work \citep{kingfield.picca_2018} and uses $Z_{DR}$ anomalies instead of fixed thresholds, which reduces the influence of the $Z_{DR}$-Bias on the $Z_{DR}C$ detection. They used thresholds of 1 or 2 standard deviations of $Z_{DR}$ from the mean of all the $Z_{DR}$ values measured at each elevation angle and then identified contiguous areas of these deviations in the elevation scan closest to 1 km above the 0°C level. A related approach, the "Hotspot-Technique", was recently introduced by \citet{krause.klaus_2024}. They calculate localised $Z_{DR}$ anomaly values based on a CAPPI at -10°C height. The localised anomaly approach is less susceptible to errors due to differential attenuation compared to the TRENDSS approach.

Most of the $Z_{DR}C$ studies using automated detection were performed in the US using S-band radar data. Some notable exceptions are the studies by \citet{plummer.etal_2018}, who looked at shallow convection using an adapted SN15 on X-band radar data, \citet{schmidt_2020}, who investigated $Z_{DR}C$ characteristics in C-band data in Germany (also using an adapted SN15), and \citet{lo.etal_2024} who implement a two-step $Z_{DR}C$ detection algorithm on a C-band radar composite for the purpose of early detection of severe convection.  A more complex $Z_{DR}C$ detection algorithm has been introduced by \citet{klaus.etal_2023} in two case studies on Austrian X- and C-band radar data. 

\begin{figure*}[htb!]
\begin{center}
\includegraphics[width=1.7\columnwidth]{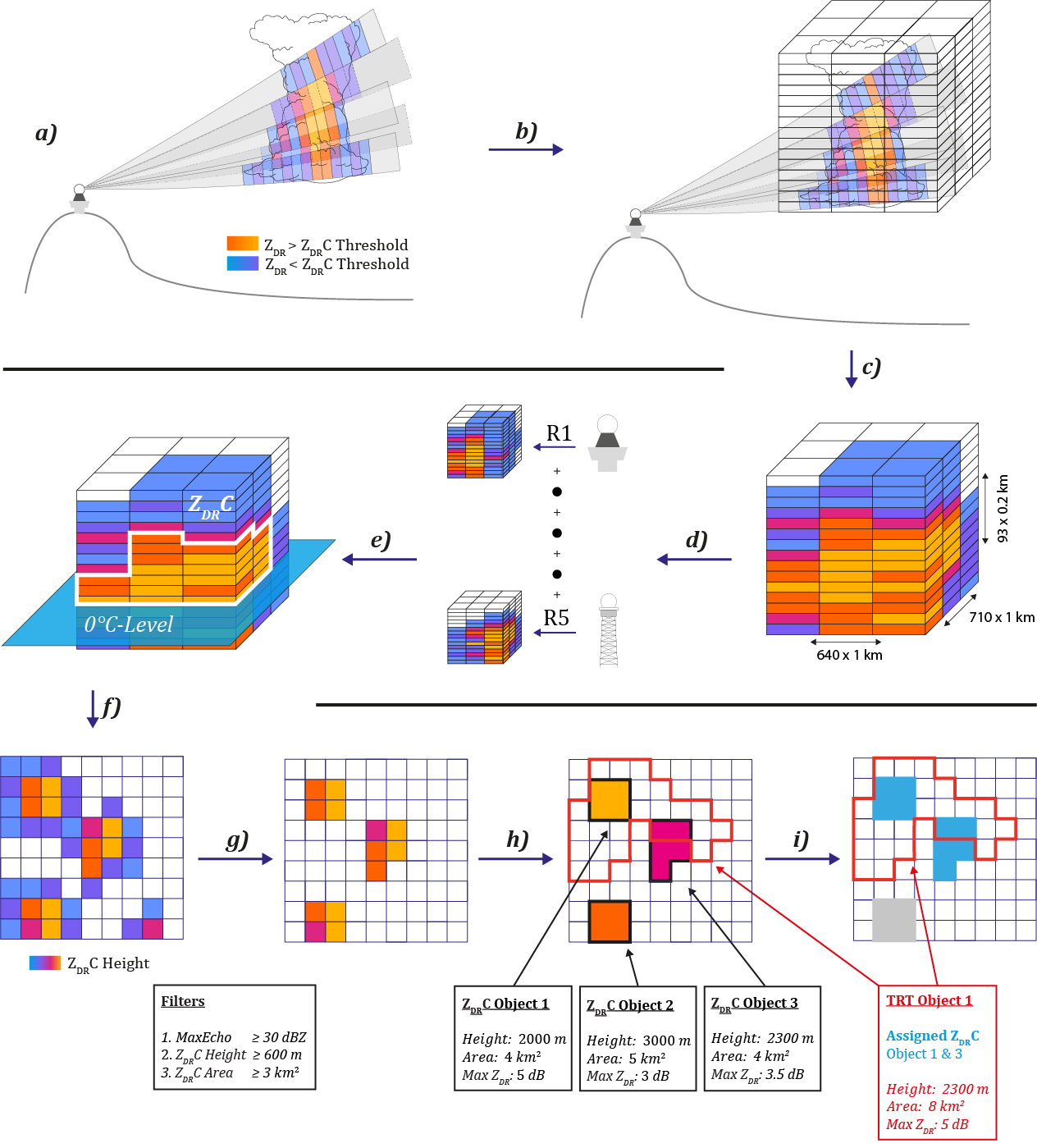} % fullwidth if possible, messes with the layout
\caption{{$Z_{DR}C$ detection algorithm steps: a) $Z_{DR}$ measured by each individual radar is bias-corrected and filtered before it is gridded onto a 3D Cartesian grid with a 1x1 km horizontal and 0.2 km vertical resolution (b,c). d) The data from all five radars are then combined into one composite dataset by taking the maximum for each pixel. e) The $Z_{DR}C$ are detected following the SN15 approach. Starting from the freezing level height, the number of contiguous layers with $Z_{DR}$ above the detection threshold for each vertical column are counted, resulting in a 2D-grid of $Z_{DR}C$ heights (f). g) To reduce noisiness, the $Z_{DR}C$ height grid is filtered using a MaxEcho, a $Z_{DR}C$ height, and a $Z_{DR}C$ area threshold. h) Individual $Z_{DR}C$ objects are detected, and their characteristics (area, height, ...) are computed. i) Finally, the $Z_{DR}C$ objects are assigned to TRT storm objects. Any $Z_{DR}C$ object intersecting a TRT object is assigned to said TRT object and the characteristics are aggregated. Areas and Volumes are summed up, and the maximum values of the other characteristics are calculated based on the combined present $Z_{DR}C$ objects.
{\label{Method_diagram}}%
}}
\end{center}
\end{figure*}

In this study, we use an adapted version of the SN15 due to ease of implementation and comparability to the study by \citet{schmidt_2020}, who also investigated the links between $Z_{DR}C$ characteristics and hail with C-band radar data. 
Our implementation of the SN15 uses the data from all five Swiss radars to create a Cartesian $Z_{DR}$ composite over the study area.  The $Z_{DR}C$ are detected on the Cartesian composite. The steps of the algorithm are shown schematically in Fig. \ref{Method_diagram}.

Before the creation of the composite, the data from each radar is corrected separately for $Z_{DR}$ biases. This correction is done following the approach introduced by \citet{dixon.etal_2017} with slight modifications. We use the operational hydrometeor classification algorithm “Hydroclass” developed by \citet{besic.etal_2018} to identify radar gates containing dry snow. The $Z_{DR}$ values of these gates are then collected for each day (subject to several quality constraints), and the median $Z_{DR}$ value is calculated. The deviation of this median from 0.2 dB (expected $Z_{DR}$ value of dry snow \citep{dixon.etal_2017}) corresponds to the $Z_{DR}$ bias of that day \citep{helmus.collis_2016, ventura.etal_2020}. If this $Z_{DR}$ bias exceeds 0.2 dB, the radar data is corrected by subtracting the calculated bias value from the absolute $Z_{DR}$ values.

The corrected data is then further filtered using a $\rho_{HV}$ minimum threshold of 0.8 to remove potential TBSS and non-meteorological signatures, and a 3-range-gate-moving-average filter is applied to reduce noisiness. A sensitivity analysis of $\rho_{HV}$ thresholds between 0.5 and 0.9 was performed, and it was found that 0.8, used in SN15, also proved adequate for the C-band when using a multi-radar composite with a relatively coarse resolution.  Comparison studies \citep{kaltenboeck.ryzhkov_2013} have shown that in C-band $\rho_{HV}$  can have lower values in hail than in S-band, with values dropping to 0.6-0.7 in the hail area. Consequently, our threshold of 0.8 might wrongly filter out some hail areas from the $Z_{DR}C$. However, interpolating the radar data to the $1 km^2$ grid counteracts this issue. For $Z_{DR}C$ detection at higher resolution, a lower $\rho_{HV}$ threshold may be necessary. 

The radar composite is computed by first gridding the polar data of each radar onto a separate Cartesian grid with a 1x1 km horizontal and 0.2 km vertical grid resolution. The gridding is done by assigning the highest measured $Z_{DR}$ value within each Cartesian grid cell as its value  (Fig. \ref{Method_diagram}, steps a and b). All five grids are then combined by simply using the maximum value measured at each grid point (Fig. \ref{Method_diagram}, steps c-e). The compositing is done for two main reasons: Firstly, we aim to have a continuous product for the entire study area with the best possible spatial and temporal coverage. This also allows the $Z_{DR}C$ detection to work if individual radars are unavailable due to maintenance work leading to a more robust algorithm. Secondly, compositing is a simple way to reduce the effects of differential attenuation without implementing an attenuation correction scheme. 

\begin{figure*}[hb!]
\begin{center}
\includegraphics[width=2\columnwidth]{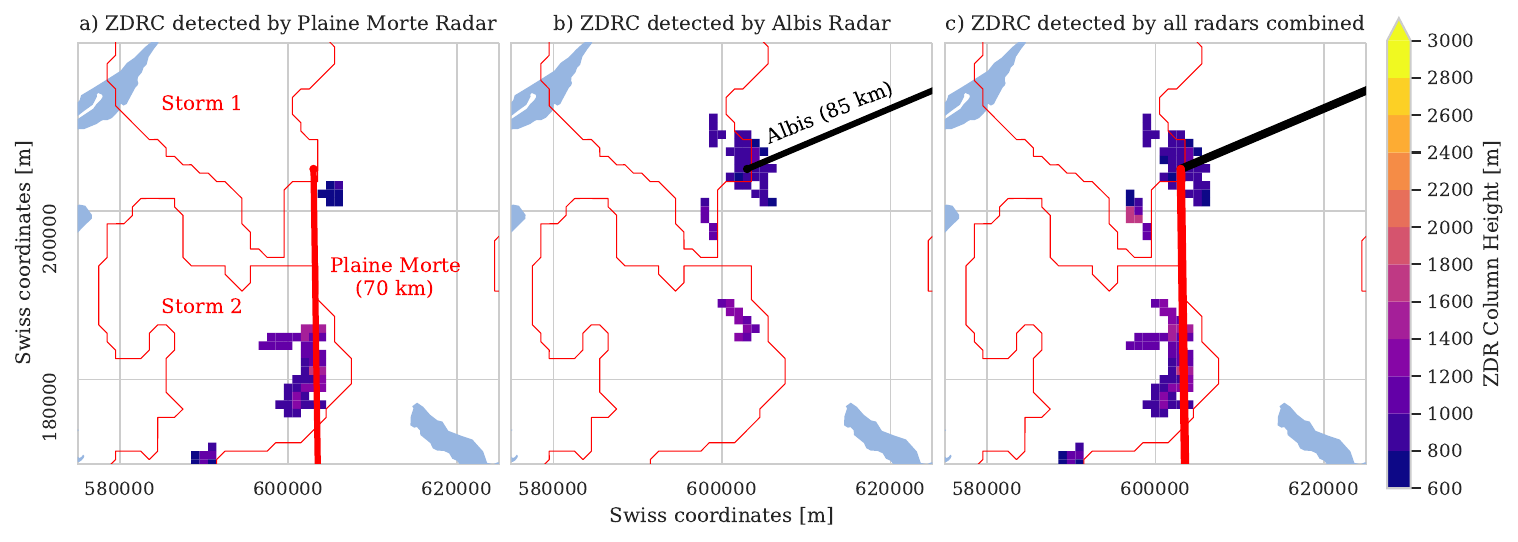}
\caption{{Example case showing the advantage of compositing multiple radars. The $Z_{DR}C$ heights (colour shading) are shown for two storms (thin red contours) over central Switzerland. a) The line of sight (thick red line) of the closest radar (Pointe de la Plaine Morte) is blocked by the convective storm closer to the radar, leading to limited detection of the storm further north. b) The Albis radar, located at a greater distance has an unobstructed view (thick black line) on the northern storm but does not detect the full extent of the southern storm. c) Combining both radars leads to better detection of the $Z_{DR}C$ in both storms.
{\label{differential attenuation}}%
}}
\end{center}
\end{figure*}

Differential attenuation at C-band in large rain and melting hail can reach values between $0.2{-}2 \, \text{dB/km}$ \citep{borowska.etal_2011}, which is sufficiently high to impact $Z_{DR}C$ detection. An example is shown in Fig. \ref{differential attenuation}, where the attenuation caused by one convective cell “blocks” the detection of the $Z_{DR}C$ of another convective cell from the closest radar (Pointe de la Plaine Morte). Most of this $Z_{DR}C$ only becomes visible using a second radar (Albis). 

Once the $Z_{DR}$ composite is computed, the $Z_{DR}C$ are calculated following SN15 using $H_0$ obtained from the COSMO-1E analysis (specifics of $H_0$ computation in \citet{trefalt.etal_2022}, chapter 3.2.2). Multiple $Z_{DR}$ thresholds between 1 and 2.5 dB were tested on 10 case study days, and it was found that 2 dB performed best for identifying $Z_{DR}C$ associated with hail-producing convective cells. Lower thresholds led to the detection of more $Z_{DR}C$ but also increased false detections (Fig. \ref{Method_diagram}, step f).

Finally, the resulting 2D field of $Z_{DR}C$ heights is filtered using a MaxEcho minimum reflectivity threshold of 30 dBZ. The reflectivity filter removes $Z_{DR}C$ artefacts outside convective storms (e.g., melting layer signatures). The $Z_{DR}C$ are then identified as contiguous objects, and all $Z_{DR}C$ with an area smaller than $3\, \text{km}^2$ and a maximum height lower than 600 m are discarded to reduce noisiness. For the final dataset of $Z_{DR}C$-objects, the $Z_{DR}C$ characteristics maximum height, maximum $Z_{DR}$ value within the column, volume, and areal extent are computed (Fig. \ref{Method_diagram}, steps g and h). Areal extent here refers to the area covered by the $Z_{DR}C$ at $H_0$. 

\subsection{Linking TRT-Cells, \texorpdfstring{$Z_{DR}C$}{ZDRC}, POH and MESHS} 

To link $Z_{DR}C$ characteristics to the TRT-cells, any $Z_{DR}C$-object whose polygon intersects with the TRT-cell’s polygon is assigned to said TRT-cell. $Z_{DR}C$ objects that do not intersect with a TRT-cell are discarded for the purpose of this study. The $Z_{DR}C$ characteristics are then assigned to the TRT-cell. If there are multiple $Z_{DR}C$ assigned to the same TRT-cell, their characteristics are aggregated; the areas and volumes are summed up, and the $Z_{DR}C$ height maximum, as well as the maximum $Z_{DR}$ value within the $Z_{DR}C$, are calculated over the combined area (Fig. \ref{Method_diagram}, step i).
For both MESHS and POH, the maximum value reached by three contiguous pixels within the TRT-polygon outline for each time step is calculated and assigned to the storm as its respective characteristic value.

\subsection{Linking TRT Cells with Hail reports and No-Hail TRT-Cells}

Thanks to the large number of available hail reports, we can use a relatively strict approach to link them to the TRT-cells.  For every cell, in each 5-minute time step, all reports made in the following 10 minutes within the boundary of the TRT-cell polygon are assigned to said polygon. With this method, there are 8346 cells with both a detected $Z_{DR}C$ and at least one hail report. 

To improve the likelihood of capturing actual hail-producing cells, we further filter these cells. The TRT-cell is assigned the largest hail size category, for which there are at least \textbf{three} reports within a 10-minute window. Using this strategy, we obtain a dataset of 4463 TRT cells with hail confirmed and a $Z_{DR}C$ during their lifetime. 2345 in the “small” category, 1127 in “medium”, and “287” in “large”. The cells in the different categories sum up to fewer than the total number of hail cells because we specifically require there to be no single report larger than the category at any point during the cell's lifetime. The aim of this study is to quantify the ZDRC characteristics related to specific hail sizes. If we were to allow single larger reports, it would introduce further uncertainty in regards to the largest actually observed hail size.

Furthermore, we also attempt to define “No-Hail” storms for comparison. The crowdsourced reports are most likely to be reported in populated areas \citep{barras.etal_2019}. In Switzerland, the MeteoSwiss weather app (and its hail reporting function) has reached sufficient widespread use that we can expect hail to be reported with reasonable certainty if the population density is high enough. Comparisons between the radar hail probability product POH and crowdsourced reports have shown that this population threshold is at a density of roughly 100 individuals per $km^2$ in recent years \citep{kopp.etal_2024}. 

We now define the No-Hail storms as storms where the median population density in a $2 \, km$ radius around the TRT-cell centroid is at least $100 \,p/km^2$ for the entire lifetime of said TRT-cell. Additionally, the dataset is limited to storms that occur between 6:00 and 23:00 local time when people are awake and will report. With this definition, we find 1621 “No-Hail” cells, 929 of them linked to a $Z_{DR}C$. The cells are, by selection, concentrated in the most populated areas of Switzerland (Fig. \ref{study area}b) and mostly short-lived due to the requirement that they must spend their entire lifetime over populated areas. It follows that they are most likely not a representative sample of all non-hail-producing storms in Switzerland, and for all comparisons to the hail-producing storms, this should be kept in mind. However, this is currently our most promising approach to “confirm” a lack of hailfall. 
In the subsequent chapters, we will use the following terminology to refer to storms (TRT-objects) with different hail size categories: No-Hail-Storms (NoHS), (any) Hail-Storms (AllHS), Small/Medium/Large Hail storms (SmHS/MedHS/LgHS).

\section{Results}

Using the 2 dB $Z_{DR}$ threshold, $Z_{DR}C$s are identified in 54\% of all studied storms (29'343) at some point in their lifetime. Lower $Z_{DR}$ thresholds lead to a larger fraction of storms containing $Z_{DR}C$s; for example, at 1.5 dB, $Z_{DR}C$s are found in 64.7\% of all storms. These additionally detected $Z_{DR}C$s tend to fall into two categories: they are either very small and shallow $Z_{DR}C$ and associated with smaller and weaker convective cells or they are shallow $Z_{DR}C$s with a very large spatial extent. These larger $Z_{DR}C$s are mainly detected close to the radar locations and are potentially a mix of melting layer signatures and widespread shallow convection. Since shallow convection is unlikely to produce updrafts of sufficient strength to produce large hail, we accept potential misses of this type of convection and use a higher 2 dB $Z_{DR}C$ threshold, reducing the total number of detected $Z_{DR}C$ but increasing the robustness of the detection.
Focusing on the confirmed hail-producing storms only, $Z_{DR}C$ were found in 81\% of all AllHS, with higher percentages at larger hail sizes, 74\% for SmHS, 93\% for MedHS, and 95\% for LgHS cases. NoHS storms are associated with $Z_{DR}C$s in 61\% of cases.

\subsection{Lifetime Maxima of the \texorpdfstring{$Z_{DR}C$}{ZDRC} Characteristics}

The first step in characterising the $Z_{DR}C$ is to look at the typical values and defining characteristics during their lifetime. Fig. \ref{lifetime boxplots} shows lifetime maximum values for the different $Z_{DR}C$ characteristics separated by storm type.

The $Z_{DR}C$ characteristics (areal extent, maximum $Z_{DR}$ within the $Z_{DR}C$, maximum $Z_{DR}C$ height, and $Z_{DR}C$ volume) differ significantly between AllHS and NoHS. Further, their distributions also differ significantly between the different hail-size classes. The differences are significant (Kolmogorov-Smirnov tests, p<0.05) between all storm types for all characteristics. The MedHS and LgHS show very similar distributions in their characteristic values. For this reason, they will be aggregated into one "Severe Hail Storm" (SevHS) class in several of the following chapters. 
In general, all hail storms show larger values in all investigated $Z_{DR}C$ characteristics than NoHS and the bigger the reported size category, the larger the values. 

\subsubsection{Characteristic Values}

\begin{figure*}[ht!]
\begin{center}
\includegraphics[width=2\columnwidth]{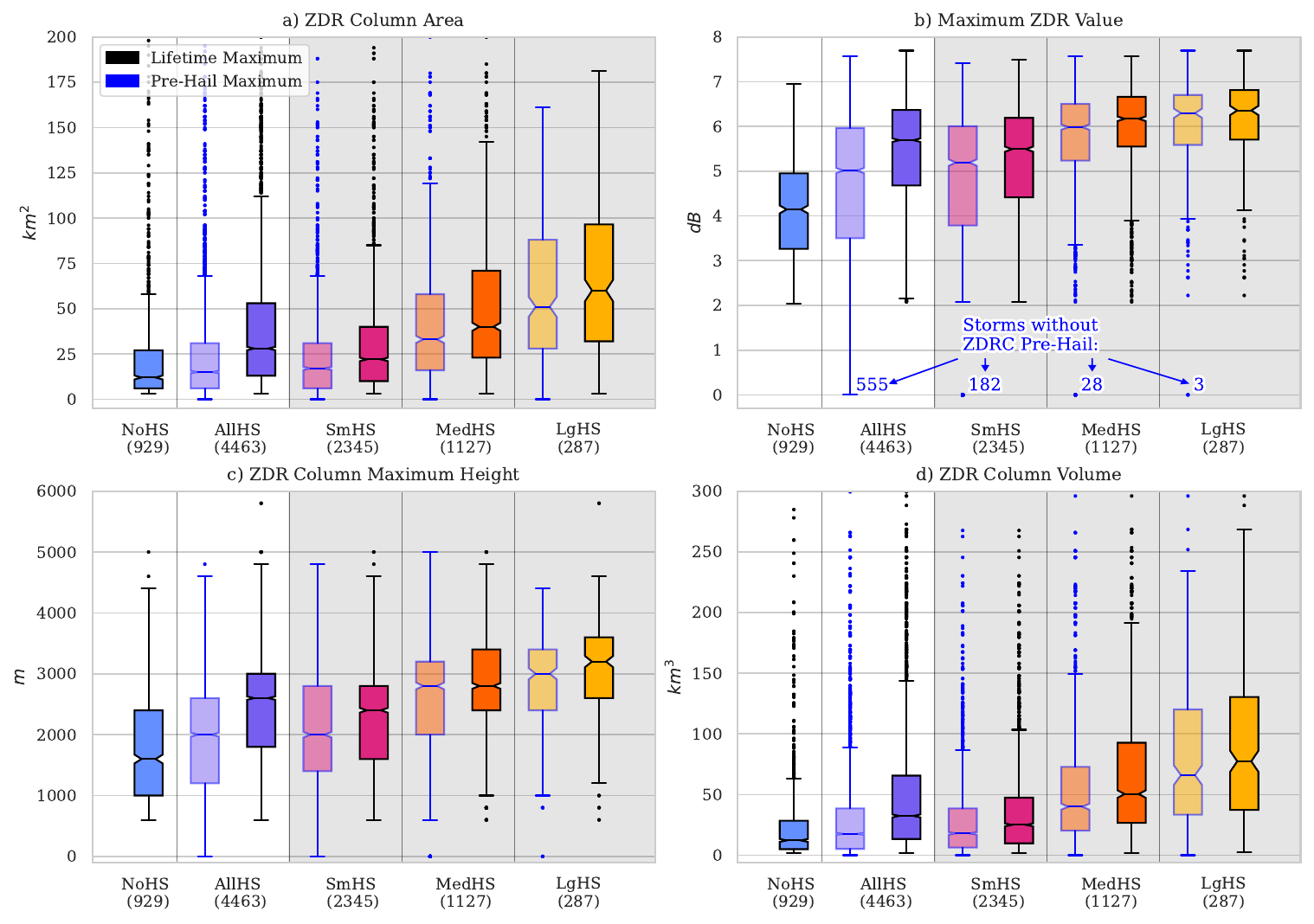}
\caption{{Box plots showing the distribution of the maximum $Z_{DR}C$ characteristic values over the entire lifetime of the storms (black outline, dark tint) and before the first hail reports of the respective categories (blue outline, light tint). The storms are separated into hail-size categories. The numbers in brackets under the category name show the number of storms in each category. The blue numbers in subplot b show the number of storms without $Z_{DR}C$ before the first hail reports.
{\label{lifetime boxplots}}%
}}
\end{center}
\end{figure*}

\textbf{$Z_{DR}C$ Area:} Typical values (defined here as values in the $25^{th}-75^{th}$ percentile range) of lifetime maximum $Z_{DR}C$ area (Fig. \ref{lifetime boxplots}a and Fig. \ref{maximum_value_distribution}a) are $6-27 \,km^2$ for NoHS  and $13-53 \, km^2$ for AllHS. The high number of SmHS compared to the other size categories largely dominates the AllHS values.
SmHS exhibit maximum areas between $10-40 \, km^2$. Looking solely at the larger hail categories, typical areas are $23-71 \, km^2$ for MedHS and $32-96 \, km^2$ for LgHS.\\\\
\textbf{Maximum $Z_{DR}$ value within the $Z_{DR}C$:} The strongest separation between NoHS and all hailstorm classes can be seen when looking at the maximum $Z_{DR}$ value within the $Z_{DR}C$ (Fig. \ref{lifetime boxplots}b and Fig. \ref{maximum_value_distribution}b). NoHS show typical lifetime maximum $Z_{DR}$ values between $3.3-5 \, dB$. In contrast, for AllHS the values are $4.7-6.4 \, dB$, for SmHS $4.4-6.2 \, dB$, for MedHS $5.6-6.7 \, dB$ and for LgHS $5.6-6.8 \, dB$. \\\\
\textbf{$Z_{DR}C$ Height:} $Z_{DR}C$ height shows a similar pattern to the previous characteristics (Fig. \ref{lifetime boxplots}c and Fig. \ref{maximum_value_distribution}c). Typical lifetime maximum $Z_{DR}C$ maximum heights for NoHS are $1000-2400 \, m$ versus $1800-3000 \, m$ for AllHS. For SmHS maximum heights typically reach $1600-2800 \, m$, for MedHS $2400-3400 \, m$ and for LgHS $2600-3600 \, m$ (Fig. \ref{lifetime boxplots}c). \\\\
\textbf{$Z_{DR}C$ Volume:} Finally, the $Z_{DR}C$ volume also shows larger values for the hailstorm categories: $5-28 \, km^3$ for NoHS compared to $13-66 \, km^3$ for AllHS, $10-47 \, km^3$ for SmHS, $27-93 \, km^3$ for MedHS, and $37-130 \, km^3$ for LgHS.
Both $Z_{DR}C$ volume and area show many high outliers, which implies a strong positive skew of the distribution. This can also be seen in Fig. \ref{maximum_value_distribution}d.

\subsubsection{Using lifetime maxima to classify storm types}

\begin{figure*}[ht!]
\begin{center}
\includegraphics[width=2\columnwidth]{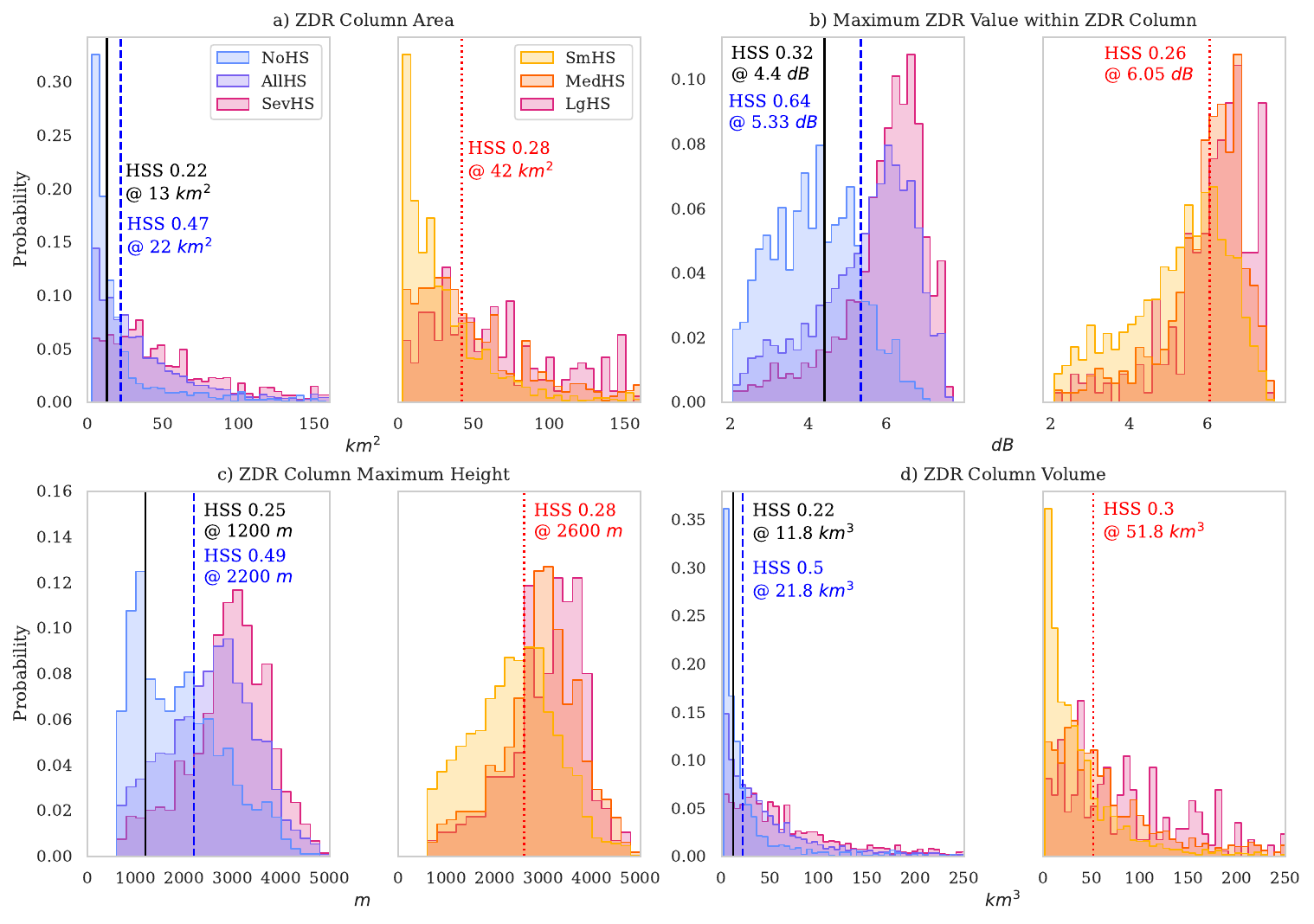}
\caption{{Histograms showing the distribution of the lifetime maximum values for each $Z_{DR}C$ characteristic separated by storm category. Further, the threshold values with the best HSS for separating between NoHS \& AllHS (black line), between NoHS \& SevHS (blue dashed line), and between SmHS \& SevHS (red dotted line) are indicated for each characteristic.
{\label{maximum_value_distribution}}%
}}
\end{center}
\end{figure*}

While the distributions of the characteristics of the different storm categories differ significantly, they have substantial overlap, as can be seen in Fig. \ref{maximum_value_distribution}, making it difficult to use them for classification purposes. Using a threshold based on a single variable has limited skill. In Fig. \ref{maximum_value_distribution}, the best-performing threshold values to separate NoHS from AllHS (solid black line), NoHS from SevHS (dashed blue line), and SmHS from SevHS (dotted red line) are shown. To assess the skill, the Heidke-Skill-Score is used (HSS).

To differentiate between NoHS and hail-producing storms (AllHS), the maximum $Z_{DR}$ value within the column is the single variable that performs best. NoHS vs AllHS can be separated with a skill of $0.32$ when using a $Z_{DR}$ threshold of $4.4 \, dB$. NoHS vs SevHs are the most clearly distinguishable storm types with a skill of $0.64$ when using a $Z_{DR}$ threshold of $5.3 \, dB$. However, to distinguish between the hailstone size categories, the maximum $Z_{DR}$ value performs worst with a skill of $0.26$ when using a threshold of $6.05 \, dB$. To distinguish hailstone size categories, the other characteristics perform slightly better, with the $Z_{DR}C$ Volume showing the best result with a skill of 0.3 when using a volume threshold of $51.8 \, km^3$. 
Overall, all characteristics show the most skill in separating SevHS from NoHS and the least in separating AllHS from NoHS.

\subsubsection{Comparison to MESHS \& POH}

\begin{figure*}[htb!]
\begin{center}
\includegraphics[width=2\columnwidth]{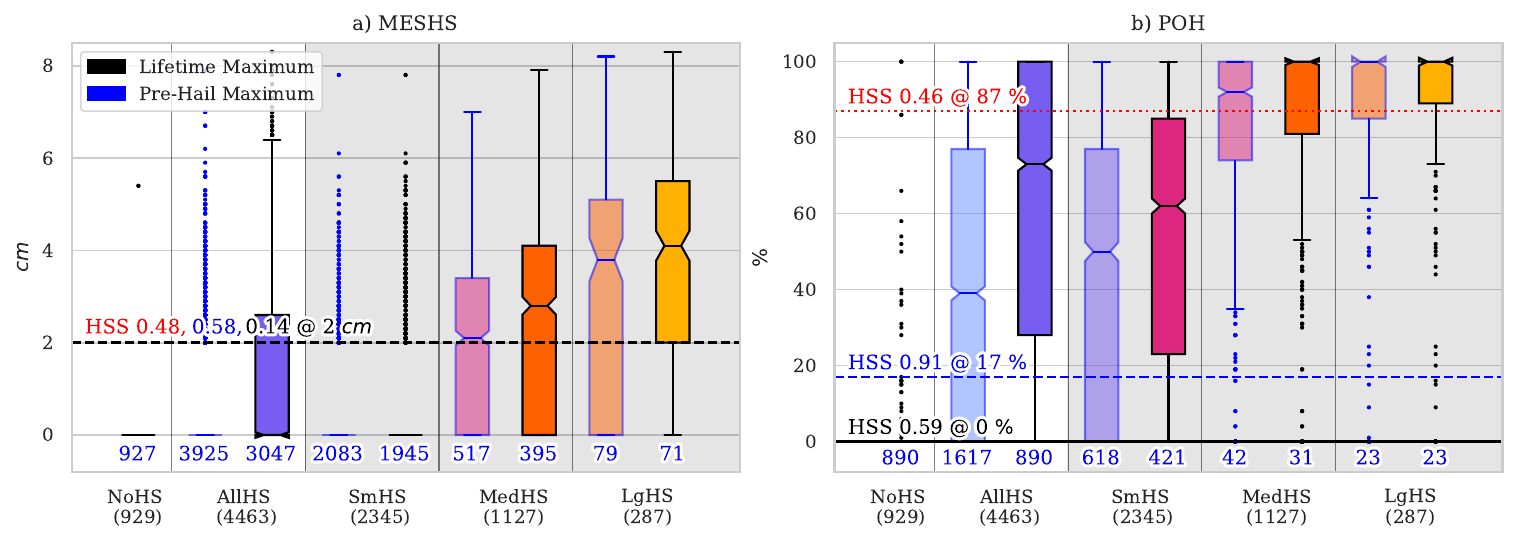}
\caption{{Same as Fig. \ref{lifetime boxplots} but for MESHS (a) and POH (b). Here, the blue numbers at the bottom of the plot show the number of storms with no MESHS/POH values measured. Further, the threshold values with the best HSS for separating the different storm types are shown as in Fig. \ref{maximum_value_distribution}.
{\label{MESHS_POH_Performance}}%
}}
\end{center}
\end{figure*}

In Fig. \ref{MESHS_POH_Performance} the currently operational radar hail products MESHS (a) and POH (b) are analysed with the same object-based approach as the $Z_{DR}C$ in the previous two chapters for comparison purposes. 

MESHS is only defined starting at $2\,cm$, which is also the best performing "threshold". Using "any" MESHS measurement as a discriminator between NoHS and AllHS results in an HSS of 0.14, which is worse than the best $Z_{DR}C$-characteristic-only-based approach (HSS 0.32, Maximum $Z_{DR}$). Similarly, for NoHS vs SevHS, the HSS is 0.58, which is slightly lower than the $Z_{DR}C$ approach (HSS 0.64, Maximum $Z_{DR}$). For SmHS vs MedHS, however, MESHS performs better (HSS 0.48) than the $Z_{DR}C$ approach (HSS 0.3, $Z_{DR}C$ Volume).

The second operational radar hail product, POH, is designed to detect hail presence but is not intended to discriminate between different hail size categories. Nevertheless, here, it performs decidedly the best, outperforming both the $Z_{DR}C$ and MESHS-based approach in discriminating NoHS vs AllHS (HSS 0.59), NoHS vs SevHS (HSS 0.91) and SmHS vs SevHS (HSS 0.46).

\subsection{Timing of \texorpdfstring{$Z_{DR}C$}{ZDRC} peak values}

\begin{figure*}[htb!]
\begin{center}
\includegraphics[width=2\columnwidth]{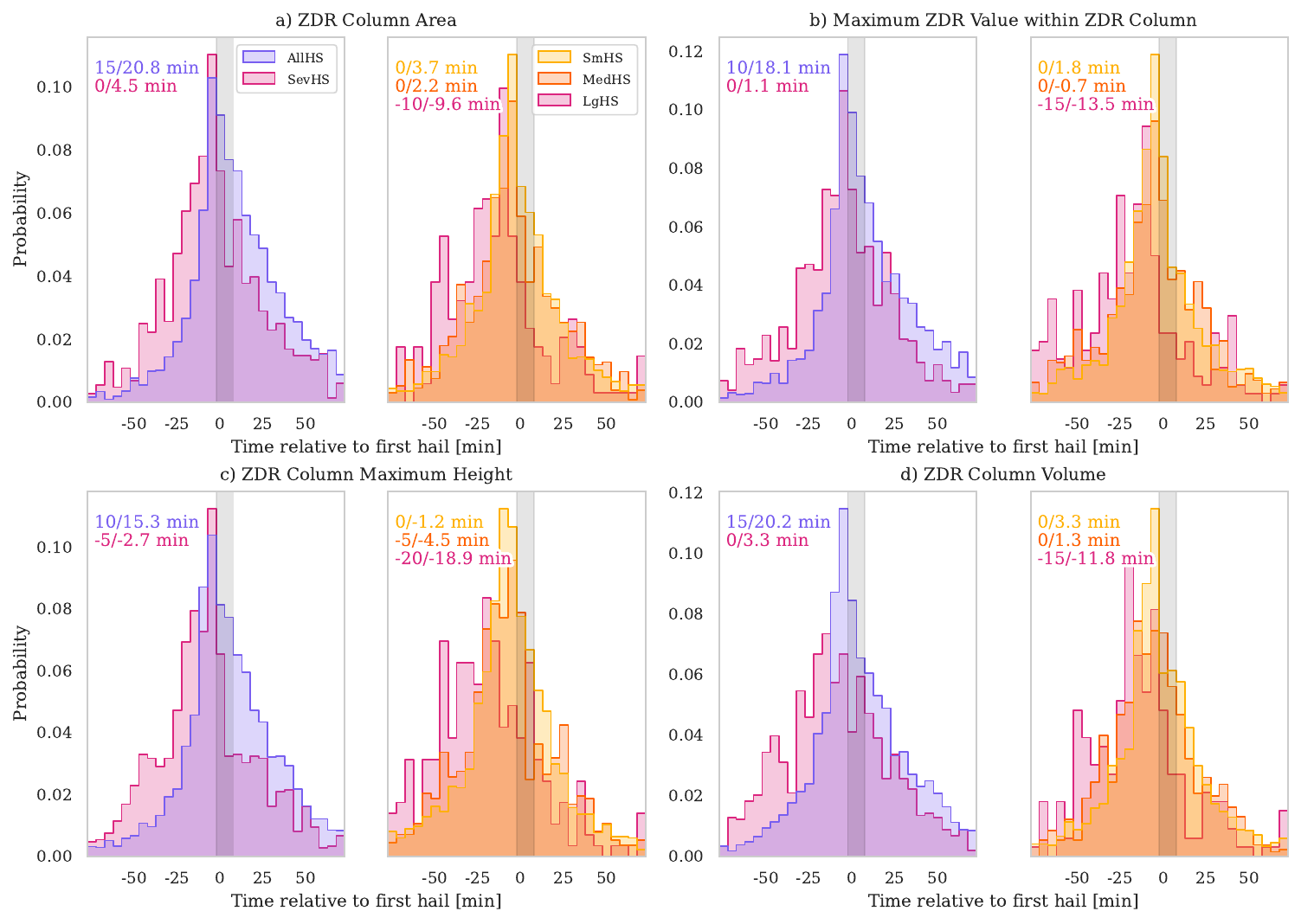}
\caption{{Histograms showing the timing of the peak values measured for each $Z_{DR}C$ characteristic variable for each storm relative to the first hail reports of the respective size category. The grey area shows the time frame during which the hail reports for determination of the first hail time-step have been made. The numbers in the panels show the median/mean values for each storm type.
{\label{peak timing 10min}}%
}}
\end{center}
\end{figure*}

In the previous section, we discuss the maximum values that the $Z_{DR}C$ exhibits throughout its lifetime in various characteristics. To assess how useful potential $Z_{DR}C$ threshold-based approaches are for hail nowcasting and real-time hail warnings, we must investigate the timing of the $Z_{DR}C$ in relation to the hail fall. 
In addition to the lifetime maxima, Fig. \ref{lifetime boxplots} also shows the maxima of each $Z_{DR}C$ characteristic before the first confirmed hail report. All $Z_{DR}C$ characteristics show slightly lower pre-hail maximum characteristic values compared to the lifetime maxima. This is partially explained by the fact that not all storms exhibit a $Z_{DR}C$ before the first hail reports (Fig. \ref{lifetime boxplots}b, blue numbers). However, even excluding these storms, the difference between the lifetime and pre-hail maxima remains similar (not shown). Consequently, this implies that not all storms exhibit the $Z_{DR}C$ maxima before the first confirmed hail reports.
 The histograms in Fig. \ref{peak timing 10min},  show the temporal distribution of the maximum measured value for each $Z_{DR}C$ characteristic relative to the first hail reports. Both the radar and the hail data are aggregated into 5-minute bins; consequently, there are uncertainties in the exact timings and the results shown here should be interpreted more as tendencies than exact values. The 10-minute aggregation window for the crowdsourced reports combined with the radar scanning time of 5 minutes means that there is an uncertainty of about 10-15 minutes in the timing. Additionally, the fall time of the hailstones adds to this uncertainty.
For all storm types and all $Z_{DR}C$ characteristics, the peak values were most frequently measured close to the first reported hail. However, all the histograms show a wide spread of more than 50 minutes before and after the hail report. 
The AllHS storms exhibit median values for all characteristics in the 10-15 minutes after first hail report bins, the SmHS and MedHS in the 5 - 0 minutes before hail bins and the LgHS in the 20 - 5 minutes before bins.  Of all the characteristics, the $Z_{DR}C$ height exhibits the earliest peaks.

\subsection{Temporal evolution of \texorpdfstring{$Z_{DR}C$}{ZDRC}}

Analysing the temporal evolution of the $Z_{DR}C$ proves challenging due to their intermittent nature. The $Z_{DR}C$s rapidly change their characteristics and frequently vanish and reemerge over brief periods. Five randomly selected example cells exhibiting this behaviour can be seen in Fig. \ref{ZDRC_state}, a-d.

\begin{figure*}[htb!]
\begin{center}
\includegraphics[width=2\columnwidth]{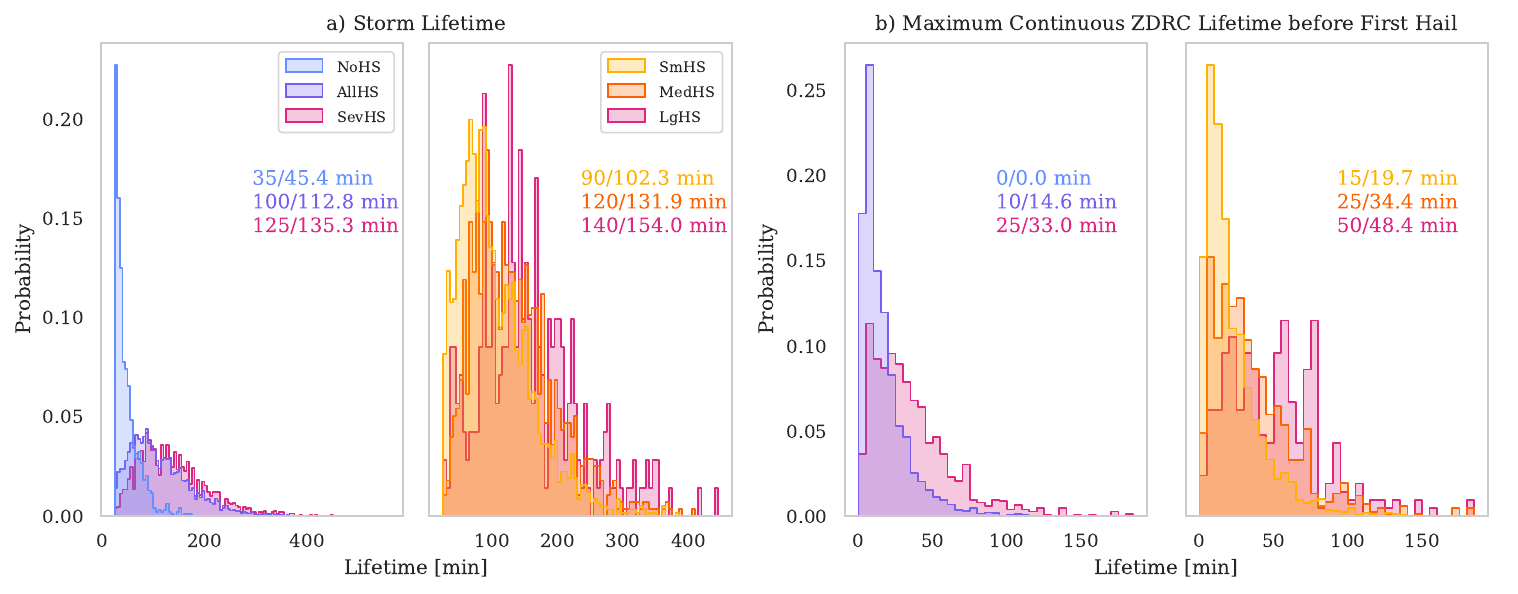}
\caption{{Histograms showing the distribution of the storm lifetimes (a) and the maximum continuous $Z_{DR}C$ lifetime before the first hail (b) for each storm category. The numbers in the panels show the median/mean values for each storm type.
{\label{maximum_ZDRC_lifetime}}%
}}
\end{center}
\end{figure*}

\begin{figure*}[htb!]
\begin{center}
\includegraphics[width=2\columnwidth]{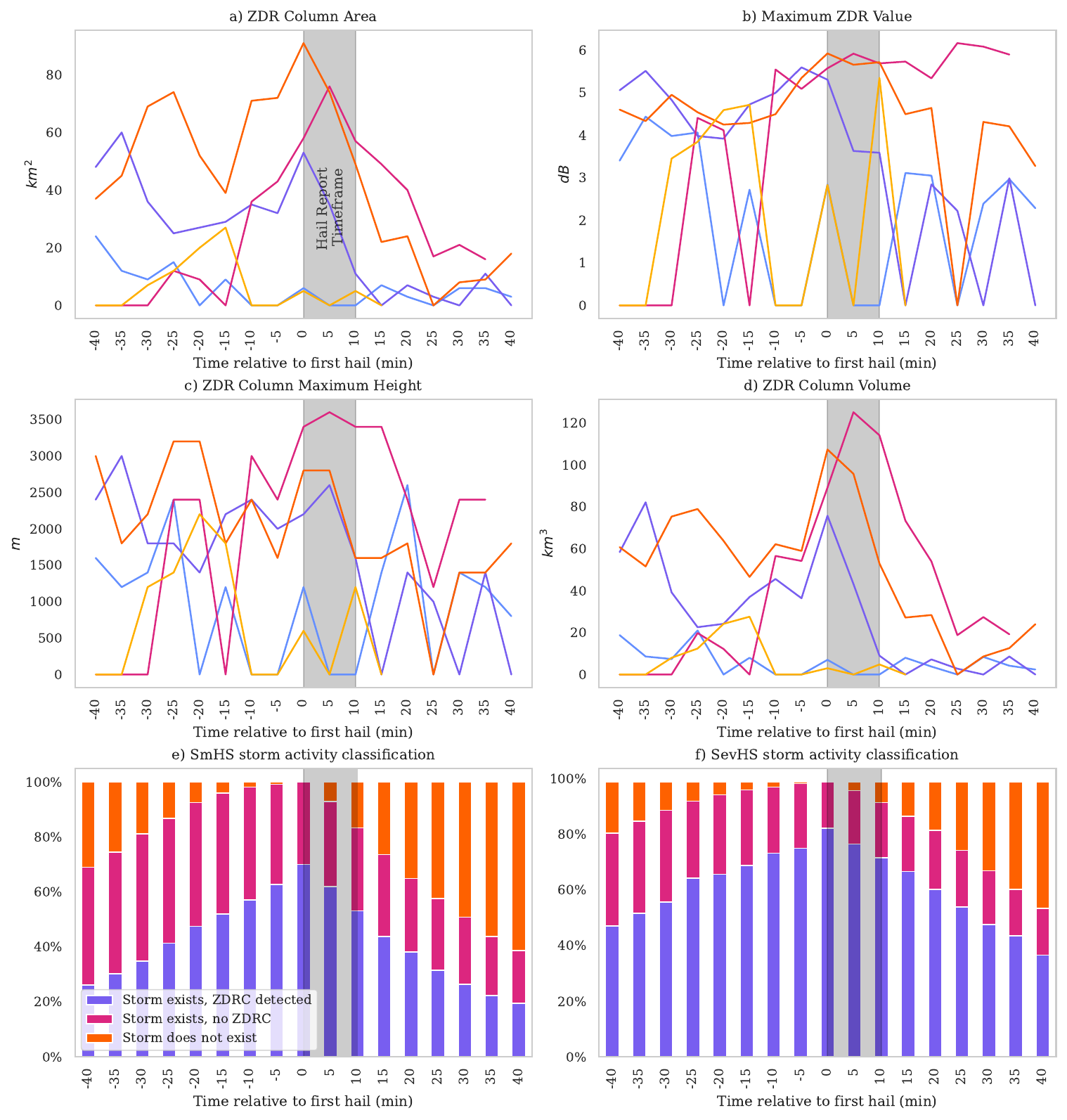}
\caption{{Subplots a-d show the different $Z_{DR}C$ characteristics relative to the time of first hail for five example storms in each category. Subplots e \& f show for SmHS (e) and SevHS (f) the fraction of total storms in different "states" relative to the first hail. If a storm is already/still detected during a specific time step, it will be categorised into "Storm exists", then depending if there is a $Z_{DR}C$ detected, it is further classified into "$Z_{DR}C$ detected" or "No $Z_{DR}C$".
The grey area shows the time frame in which the hail reports have been made to determine the first hail time-step.
{\label{ZDRC_state}}%
}}
\end{center}
\end{figure*}

In the 20 minutes preceding the first hail report, 47\% of AllHS exhibit a continuous $Z_{DR}C$. This fraction is size-dependent; it's 37\% for SmHS, 57\% for MedHS, and 61\% for LgHS. Similarly, the median continuous $Z_{DR}C$ lifetime before the first hail reports is also size dependent: It is 10 minutes for AllHS, 15 minutes for SmHS, 25 minutes for MedHS, and 50 minutes for LgHS (Fig. \ref{maximum_ZDRC_lifetime} b). This behaviour is also partly influenced by the lifetimes of the different storm types. The bigger hail categories are produced by storms with longer overall lifetimes (Fig. \ref{maximum_ZDRC_lifetime} a). The median lifetime of SmHS is 90 minutes, 120 minutes for MedHS, and 140 minutes for LgHS.

A fraction of AllHS, 18\%, do not exhibit a $Z_{DR}C$ before the first hailfall. This fraction is lower for SmHS (11\%), MedHS (4\%), and LgHS (2\%). Of the hailstorms with no $Z_{DR}C$ before the first hailfall, 45\% show a $Z_{DR}C$ during the 10-minute crowdsource collection window. While a continuous $Z_{DR}C$ can be identified in many hail storms, it is not a separating characteristic from NoHS. NoHS also show at least 20 minutes of continuous $Z_{DR}C$ presence in 45\% of cases.

The intermittent nature of the detected $Z_{DR}C$ complicates the analysis of their behaviour before hailfall. Figures \ref{ZDRC_state} e and f show the fraction of storms that exhibit active $Z_{DR}C$s at each time step relative to the first hail for each storm type. SmHS show an increase in the fraction of storms with active $Z_{DR}C$ from 26\% 40 minutes before to 70\% at the time of first hail. These values are higher for SevHS increasing from 48\% to 83\%. The number of storms which are already active 40 minutes before hail detection is lower in the SmHS at 69\% compared to SevHS 81\%, which consequentially influences the previously mentioned fraction of storms with active $Z_{DR}C$.

\begin{figure*}[htb!]
\begin{center}
\includegraphics[width=2\columnwidth]{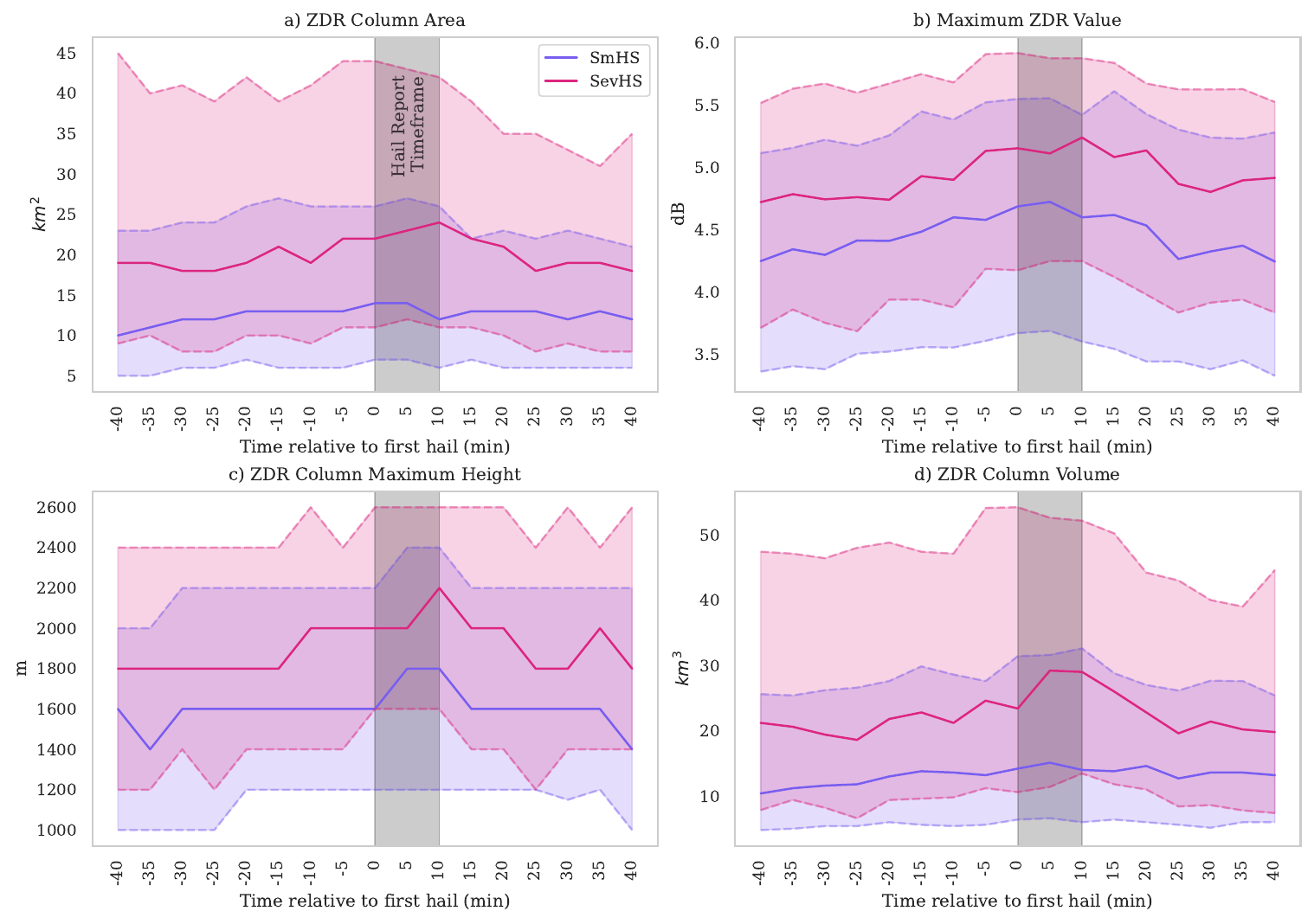}
\caption{{Same as Fig. \ref{ZDRC_state} a-d but aggregated over all active $Z_{DR}C$ at each time step. 25th percentiles (bottom dashed lines), medians (solid lines), and 75th percentiles (top dashed lines) are shown separated by storm category. 
{\label{ZDRC_relative_to_hail}}%
}}
\end{center}
\end{figure*}

Focusing only on the storms with active $Z_{DR}C$ and the development of their $Z_{DR}C$ characteristics (Fig. \ref{ZDRC_relative_to_hail}) we observe that there is a large overlap of the 25$^{th}$ to 75$^{th}$ percentile of values for the two shown storm categories for all characteristics. However, the medians significantly differ (Kolmogorov-Smirnov tests, p<0.05) between the storm categories at all timesteps for all characteristics before the first hail. 

The temporal development over the 40 minutes prior to the hail fall does not exhibit any clear trends in the $Z_{DR}C$ area, height, and volume characteristics. The maximum $Z_{DR}$ values show an increase over time until hailfall for both SmHS and SevHS before dropping again after the start of the hailfall.

\section{Discussion}

The potential of the $Z_{DR}C$ radar signature and its characteristics to be used as a tool to detect and nowcast severe storms has been described in several studies using radar observations and models. It has previously been shown both in models \citep{kumjian.etal_2014,ilotoviz.etal_2018,snyder.etal_2015} and radar-based studies \citep[e.g.][]{schmidt_2020, kuster.etal_2019,broeke_2016} that $Z_{DR}C$ characteristics can be linked to the occurrence and size of hail at the ground. Our study extends prior research by investigating $Z_{DR}C$ characteristics systematically in more than 3.5 years of operational C-band radar data and linking them to a uniquely large set of crowdsourced hail report data (173’000 reports).

The aim of this work is to investigate the potential of $Z_{DR}C$s for hail detection and nowcasting using an operational radar network in a complex orography. We implement a $Z_{DR}C$  detection algorithm based on the algorithm introduced by \citet{snyder.etal_2015} and adapt it to the Swiss C-band weather radar network. The network's high temporal and spatial resolution and the high number of radar scanning elevations make it particularly well-suited for the detection of the rapidly changing $Z_{DR}C$.  

Notably, we composite the $Z_{DR}$ data from five different C-band radars, which increases the robustness of the $Z_{DR}C$ detection, while improving spatio-temporal coverage. Specifically, we observed cases where one radar is impacted by differential attenuation, limiting its view of a storm. In these cases, overlapping radar coverage ensures adequate detection of the storm by another radar. Further, the $Z_{DR}C$ detection quality decreases with increasing distance from hail \citep{schmidt_2020}, which is also ameliorated by using multiple overlapping radars.

The detection of $Z_{DR}C$ with C-band data requires an adaptation of the thresholds used in the implementation of the algorithm at the S-band. Multiband radar studies \citep{kaltenboeck.ryzhkov_2013}, as well as theoretical simulations \citep{ryzhkov.zrnic_2019}, show that the $Z_{DR}$ of melting hail can be significantly higher at the C-band compared to the S-band. \citet{kaltenboeck.ryzhkov_2013} find $Z_{DR}$ values which are consistently above 4 dB in C-band observations compared to values below 1 dB in the corresponding S-band observations. Consequently, we performed case studies on ten days of data to determine an adequate $Z_{DR}$ threshold to identify hail-producing storms with C-band radar. We found that a 2 dB threshold performed best, which is consistent with \citet{schmidt_2020}, who performed scattering simulations at C-band, concluding that this threshold is suitable. Lowering the threshold increases the number of storms with $Z_{DR}C$s but simultaneously increases erroneous detection. For the purpose of $Z_{DR}C$ detection in deep convection, which is a prerequisite for hail production, the higher threshold, combined with various filters, proved to be adequate. Furthermore, in their modelling study on $Z_{DR}C$, \citet{kumjian.etal_2014} used a numerical model with a polarimetric C-band radar forward operator, and they found that the 2 dB $Z_{DR}$ isoline height strongly correlates with the vertical velocity of the storm's updraft at the same height.

Using the above-described detection algorithm and threshold choices, more than half of all objectively detected storms (54'624) are associated with a $Z_{DR}C$. The fraction is higher (81\%) for hail storms (4'463) and even higher (93\%) for severe hail storms (1484). These high fractions indicate that a $Z_{DR}C$ can be expected to be detected in a hailstorm. To our knowledge, no comparable analysis in the literature looks at $Z_{DR}C$ frequency in convective storms detected by an operational radar network. However, other studies that looked specifically at $Z_{DR}C$ in relation to hail reports found $Z_{DR}C$s before hail in most cases \citep{kuster.etal_2019,schmidt_2020}. Further, \citet{plummer.etal_2018} report “unexpectedly frequent” $Z_{DR}C$ in their analysis of shallow convection (not specifically hail producing) using X-band radar over the United Kingdom, though they do not quantify this frequency.

However, the $Z_{DR}C$ presence is insufficient to separate hail storms from non-hail-producing storms. In 61\% of the “confirmed” no hail storms, a $Z_{DR}C$ was also identified. The NoHS showing a higher frequency of $Z_{DR}C$ compared to AllHS could further indicate that the storms we identified as no-hail storms using our population-based definition might not be representative of all non-hail-producing storms. The selection method favours short-lived storms over the Swiss plateau, an area with relatively hilly topography compared to Switzerland in general. \\

\textit{Do the characteristics of $Z_{DR}C$ relate to observed hail on the ground?}\\

The lifetime maximum characteristics of the $Z_{DR}C$, namely $Z_{DR}C$ area, maximum height, volume, as well as the maximum $Z_{DR}$ value measured within the $Z_{DR}C$, all show significant differences between the NoHS, SmHS and SevHS, offering the potential to use these characteristics to differentiate between storm types. SevHS can be distinguished from NoHS with some skill (HSS 0.47-0.65 depending on the characteristic) when using a simple threshold approach. 

\begin{figure*}[htb!]
\begin{center}
\includegraphics[width=2\columnwidth]{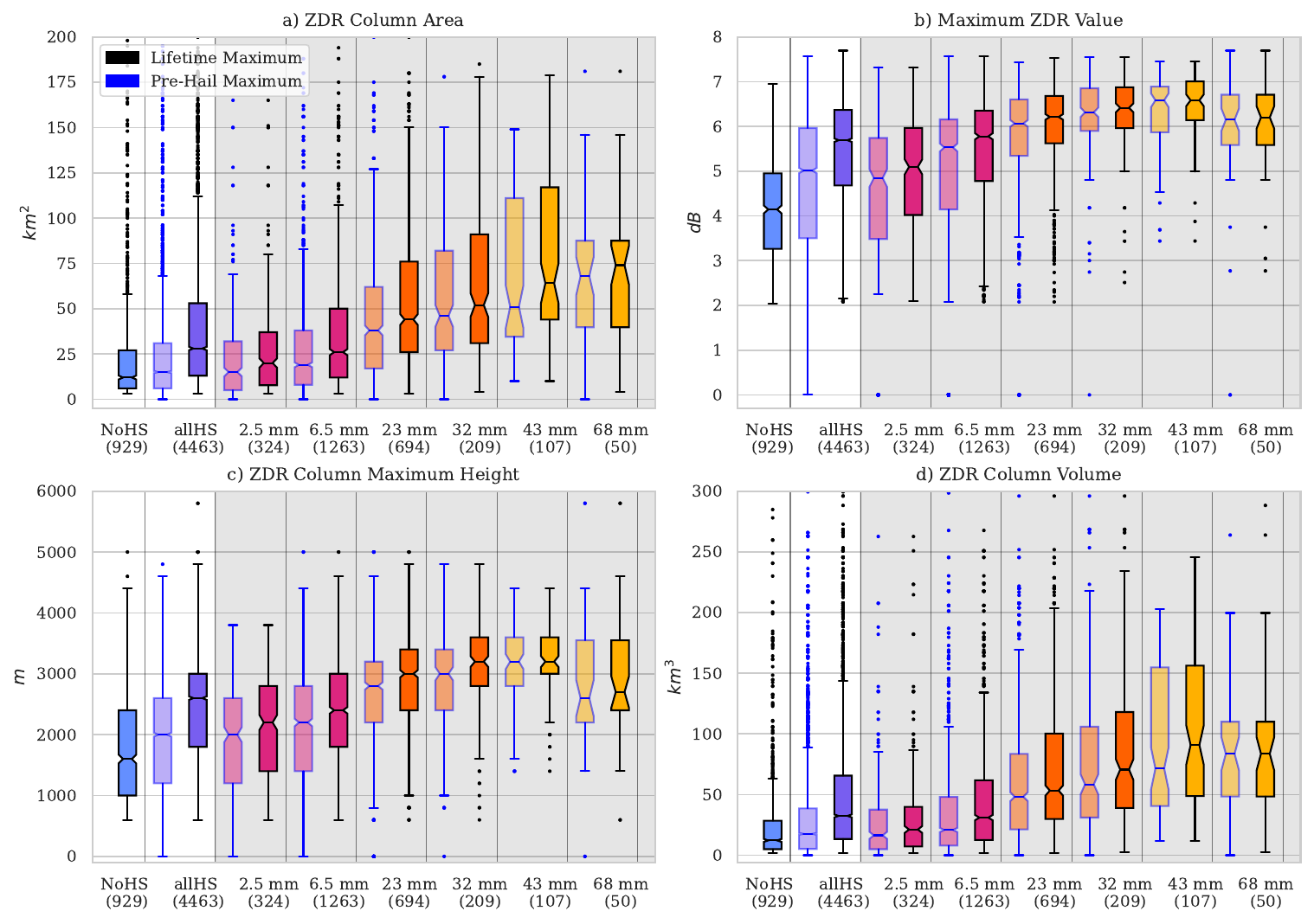}
\caption{{Same as Fig. \ref{lifetime boxplots} but with all original report categories instead of the aggregated ones. The report categories in the app are "smaller than coffee bean" (2.5 mm), "coffee bean" (6.5 mm), "one-franc coin" (23 mm), "five-franc coin" (32mm), "golf ball" (43 mm), "tennis ball" (68mm).
{\label{lifetime boxplots all categories}}%
}}
\end{center}
\end{figure*}

Distinguishing SmHS from both NoHS as well as SevHS proves to be more difficult. One cause for the vaguer distribution of $Z_{DR}C$ characteristics values for SmHS is the quality of the crowdsourcing data at small sizes. It is known that the smallest report category is often also used to report graupel instead of hail by the public \citep{barras.etal_2019}. As a result, the SmHS likely contains a mix of graupel and hail-producing storms, blurring the typical characteristics of their $Z_{DR}C$s. On the other end of the spectrum, the largest reporting category, "tennis ball", also frequently captures erroneous reports. Fig. \ref{lifetime boxplots all categories} shows the same lifetime maximum values for the different $Z_{DR}C$ characteristics as Fig. \ref{lifetime boxplots} but with the original report categories instead of the aggregated version used above. The 68 mm report category shows lower values than the 43 mm category, especially in the $Z_{DR}C$ height and the maximum $Z_{DR}$ within the $Z_{DR}C$ characteristic, which is likely due to false reports.

The typical values we measured for each $Z_{DR}C$ characteristic are comparable to the ones found in previous work \citep{plummer.etal_2018,kuster.etal_2019,wilson.broeke_2022,french.kingfield_2021, segall.etal_2022}. Maximum $Z_{DR}C$ areas for SmHS were typically (25$^{th}$ -75$^{th}$ percentile) between $10-40 \, km^2$ and between $24-76\,  km^2$ for SevHS. 

\citet{plummer.etal_2018} found varying $Z_{DR}C$ areas depending on height above freezing level, with the largest areas measured at the freezing level falling between approximately $1-12 \,  km^2$ using X-band radar. These areal extents fall on the lower side of our range; however, their study focused on shallow convection using a single radar. Conversely, using S-band radar, \citet{kuster.etal_2019} measured maximum values between approximately $7-124 \, km^2$ (10$^{th}$-90$^{th}$ percentile) for a mix of 42 storms ranging from non-severe single cells to severe supercells. These values are similar to other studies in the US focusing on tornadic storms \cite[e.g.,][]{wilson.broeke_2022,french.kingfield_2021,segall.etal_2022}.

Similarly, our values for maximum $Z_{DR}C$ heights vary between 1000-2400 m for NoHS and 2400-3400 m for SevHS, comparable to the results by \citet{kuster.etal_2019} who found values between approximately 1800-4000 m and \citet{plummer.etal_2018} who detected values between approximately 500-2500 m.  Overall, the lifetime maximum $Z_{DR}C$ height and area values are relatively similar to previous studies, even with different $Z_{DR}C$ detection algorithms, radars, and study areas.

The most distinctive characteristic of $Z_{DR}C$ to separate between storm types is the maximum $Z_{DR}$ value measured within the $Z_{DR}C$. The characteristic values of SevHS lie between 5.5-6.7 dB, which is markedly higher than the 3.3-5.0 dB of the NoHS. SmHS show values overlapping both other categories between 4.4-6.2 dB.

Comparability to other observational studies is limited as little literature looks at $Z_{DR}$ in $Z_{DR}C$, specifically using C-band data. \citet{schmidt_2020} is the exception; they investigated the same characteristic at different timesteps before hail reports and found comparable results 20-10 minutes before hail with measured values for smaller hail (1-2.5 cm) between approximately 2.2-4.5 dB and for severe hail between 4.2-7 dB. Further, studies looking at $Z_{DR}$ signatures in severe hail-producing storms independent of $Z_{DR}C$ \cite[e.g.][]{kaltenboeck.ryzhkov_2013} find high $Z_{DR}$ values of up to 6.25 dB. However, their highest values are measured below the freezing level, limiting our results' comparability.\\

\textit{Can the $Z_{DR}C$ characteristics be used to estimate the observed hail size on the ground?}\\

The $Z_{DR}C$ characteristics differ significantly between the different storm types, which allows the conclusion that there is information in the $Z_{DR}C$ that can be linked to hail size at the ground answering our first question. However, while the differences are significant, there are large overlaps in the distributions, which makes classification based solely on a single $Z_{DR}C$ characteristic difficult. A simple single-variable approach based on $Z_{DR}$ within the $Z_{DR}C$ performs slightly better than the currently operational product MESHS but worse than POH when classifying storms based on their lifetime maximum values. Potentially, an approach taking multiple $Z_{DR}C$ characteristics in to account simultaneously could perform better and should be investigated in future work.\\

\textit{Can the $Z_{DR}C$s be used to nowcast hail at the ground?}\\

The temporal evolution of the $Z_{DR}C$ in the storms of our dataset is markedly influenced by the $Z_{DR}C$'s intermittent behaviour. Only a minority of storms exhibit a continuously present $Z_{DR}C$, making the average $Z_{DR}C$ development analysis difficult and impacting the nowcasting potential. This intermittency in $Z_{DR}C$ detection has also been shown in previous studies \citep{kuster.etal_2019, schmidt_2020}. The $Z_{DR}C$ has been shown to be a rapidly evolving signature \cite[e.g.][]{kumjian.etal_2010}, meaning that the 5-minute update time of the operational radars might be too low to describe their evolution adequately. \citet{kuster.etal_2019} show how a more rapid radar volume update time influences the detection of $Z_{DR}C$ area changes. They find that a rapid-update scanning strategy (1.9 min per volume) captures the $Z_{DR}C$ signature evolution earlier than a slower update time (5.6 min), significantly influencing the prognostic possibilities. Further, the intermittency may result from the threshold-based detection of the $Z_{DR}C$. If the measured $Z_{DR}$ values fluctuate around a static threshold value, this may lead to an unstable $Z_{DR}C$ detection.

Looking solely at the timing of the above-described lifetime maximum values, we find that most storms exhibit peaks of all characteristics within 10 minutes of the reported hail at the ground. The temporal inaccuracy of crowdsourced hail reports, the 10-minute aggregation window for the reports combined with the scanning strategy employed by the radar \citep{germann.etal_2022}, and the fall of the hailstones from the cloud to the ground all contribute to the uncertainty of the time estimates. 

Notably, the AllHS storm category has later (10-15 minutes) peaks than the other storms. This is a consequence of how the timing of the first hail is defined. In the other categories, the timestamp where we first have the respective size class confirmed is used as the decisive timestamp. Consequently there may be smaller hail before in the same storm. AllHS, however, the first confirmed hail timestamp overall is used, which means there may be larger hail reported later in the storm's lifetime, which influences the $Z_{DR}C$ timing.

\citet{kuster.etal_2019} also looked at the timing of $Z_{DR}C$ height and area peaks relative to hail reports. They used data from radars in rapid update mode, providing higher temporal resolution (<2-min volumes). Nevertheless, they found that the median lead time for both characteristics was 7 minutes, which approximately matches our results. Further, \citet{kumjian.etal_2014} found that the $Z_{DR}C$ height peak correlates with increases in hail at the ground at a lag of about 10-15 minutes in their numerical model. 
We do see a slightly higher frequency of peak $Z_{DR}C$ characteristic values before the hail reports at the ground. Further, we observe that the $Z_{DR}C$ are more likely to be present before the hail than after and tend to intensify before the first hail.

The intermittency of the detected $Z_{DR}C$ limits the potential for nowcasting. An open question is how much of this intermittency is an inherent behaviour of the $Z_{DR}C$ and how much is due to limitations of the detection method. The SN15 algorithm has some inherent limitations: the fixed $Z_{DR}$ detection threshold may lead to errors in detection, specifically in areas with poor data quality, e.g., due to differential attenuation. While our composite approach mitigates this effect partially, it is still present in the data, especially along the borders of the study area, where radar coverage is limited. Further, the requirement for columns to be perfectly vertical in the algorithm may cause issues in fast-moving, tilted storms; however, we did not observe any examples where this occurred in the selected case study days.

\section{Conclusion}

The study investigates the potential use of the $Z_{DR}C$ radar signature for the detection and nowcasting of hail and its size. Previous work has shown that $Z_{DR}C$s provide information about a storm’s updraft intensity and size, which in turn influences the storm's propensity to produce hail. We use a large set of crowdsourced hail report data in combination with observations from a network of operational C-band radars to link various $Z_{DR}C$ characteristics to hail occurrence and size. Given the high density of available hail report data, we further attempt to compare the $Z_{DR}C$ characteristics in hail-producing storms to likely non-hail-producing storms. This work is distinct from prior studies in several areas: the size of the observational dataset, the study area's complex alpine orography, and the characteristics of the radars used for $Z_{DR}C$ detection.

The $Z_{DR}C$ are detected using an adapted version of the widely used algorithm introduced by \citet{snyder.etal_2015} implemented for an operational C-band radar network in complex topography.
Using this algorithm, we find that $Z_{DR}C$s are detected in the majority of hail-producing storms, with a higher likelihood of detection in more severe storms. The presence of $Z_{DR}C$ is, however, not a distinctive property of hail-producing storms, since non-hail-producing storms also show frequent $Z_{DR}C$ presence. 

Looking at the lifetime maximum values of different $Z_{DR}C$ characteristics, namely the $Z_{DR}C$ area, maximum height, volume, as well as the maximum $Z_{DR}$ value measured in the $Z_{DR}C$, we find that there are significant differences between non-hail-producing, small-hail-producing and severe-hail-producing storms. The most promising characteristic for differentiating between storm types is the maximum measured $Z_{DR}$ value within a $Z_{DR}C$.

The temporal evolution of the $Z_{DR}C$s is difficult to characterise due to their intermittent nature. The $Z_{DR}C$s undergo rapid changes in all characteristics throughout their lifetime. Generally, the peak values of all characteristics are measured before the first reported hail, potentially allowing nowcasting with some skill. There is, however, substantial variability in the timing of the peak values, limiting the nowcasting potential when using a simple threshold approach. Nevertheless, on average $Z_{DR}C$s intensify before the first hail observation, offering another avenue for potential hail nowcasting.

The results of this study show that there is potential in using $Z_{DR}C$ for the detection of hail and its size. This warrants further investigation and development of more sophisticated detection techniques. Radar data with higher temporal and spatial resolution could potentially enable the tracking of $Z_{DR}C$s over time, leading to more in-depth information about their development over time and how it influences hail generation. Further, a $Z_{DR}C$ detection approach with a dynamic detection threshold  \citep[e.g.][]{krause.klaus_2024, french.kingfield_2021, kingfield.picca_2018} should be tested to investigate whether the observed intermittent behaviour of the $Z_{DR}C$ is significantly impacted by the detection approach.

Finally, a multivariate approach using various characteristics of the $Z_{DR}C$ simultaneously, in combination with other products such as POH, should be investigated for its potential in warning applications. For example, $Z_{DR}C$ could be used as a predictor in machine learning models such as the one presented by \citet{leinonen.etal_2023} and \citet{rombeek.etal_2024}.

%\backmatter
\bmsection*{Author contributions}
\textbf{Martin Aregger}: conceptualisation; data curation; formal analysis; investigation; methodology; software; visualisation; writing – original draft; writing – review and editing. \textbf{Olivia Martius}: conceptualisation; methodology; resources; supervision; writing – review and editing. \textbf{Alessandro Hering}: conceptualisation; methodology; resources; supervision; writing – review and editing. \textbf{Urs Germann}: conceptualisation; methodology; resources; supervision; writing – review and editing.

\bmsection*{Acknowledgments}
The research leading to these results has been carried out within subproject C of the \textbf{s}eamless \textbf{c}oupling of kilometer-resolution weather predictions and \textbf{C}limate simulations with hail \textbf{im}pact assessment for multiple sectors (scClim, \hyperlink{https://scclim.ethz.ch/}) project a 4-year Sinergia project funded by the Swiss National Science Foundation (CRSII5\_201792).

\bmsection*{Data Availability Statement}
The Python code written for the calculation of the $Z_{DR}C$ is available on Github (https://github.com/MartinAregger/ZDRC\_Calculator\_QJ\_2024).
The input radar data for calculating the ZDRC and output ZDRC fields are available for a limited number of selected events upon request.

%https://github.com/jm-mustafa/Single/_Peak/_Waveforms.

\bmsection*{Conflict of interest}
The authors declare no potential conflict of interests.

\bmsection*{ORCID}
Martin Aregger: https://orcid.org/0000-0001-6741-0707 \\
Olivia Martius: https://orcid.org/0000-0002-8645-4702 \\
Urs Germann:    https://orcid.org/0000-0002-8539-7080 \\
Alessandro Hering: https://orcid.org/0000-0003-2043-3981

\bibliography{bibliography/bibliography}

%\bmsection*{Supporting information}

%\nocite{*}% Show all bib entries - both cited and uncited; comment this line to view only cited bib entries;

\end{document}